\documentclass[sigconf, screen]{acmart}

\usepackage{multirow}
\usepackage{array}
\usepackage{balance}
\usepackage{url}
\usepackage{makecell}
\usepackage{graphicx}
\newenvironment{tcolorbox}{%
  \par\smallskip
  \noindent
  \begin{minipage}{\linewidth}
  \ignorespaces
}{%
  \end{minipage}
  \par\smallskip
}

\newenvironment{examplebox}[1]{%
  \begin{quote}\small
  \noindent\textbf{#1}\par\smallskip
  \ignorespaces
}{%
  \end{quote}
}

\usepackage{fontawesome} % 添加fontawesome宏包
\usepackage{booktabs}       % Provides professional table lines (\toprule, \midrule, \bottomrule)
\usepackage{xspace}
\usepackage{color}

\newcommand{\stab}{\vspace{0.5ex}\noindent}

\newcommand{\stitle}[1]{\stab\noindent\textbf{#1}}

\newcommand{\utitle}[1]{\noindent{\it \underline{#1}}}
\newcommand{\uctitle}[1]{\textit{#1}}

\newcommand{\eg}{\textit{e.g.,}\xspace}
\newcommand{\vl}{Vega-Lite\xspace}

\AtBeginDocument{%
  }

\copyrightyear{2026}
\acmYear{2026}
\setcopyright{cc}
\setcctype{by}
\acmConference[CHI '26]{Proceedings of the 2026 CHI Conference on Human Factors in Computing Systems}{April 13--17, 2026}{Barcelona, Spain}
\acmBooktitle{Proceedings of the 2026 CHI Conference on Human Factors in Computing Systems (CHI '26), April 13--17, 2026, Barcelona, Spain}
\acmPrice{}
\acmDOI{10.1145/3772318.3791441}
\acmISBN{979-8-4007-2278-3/2026/04}
\begin{document}

\title[Debugging Defective Visualizations]{Debugging Defective Visualizations: Empirical Insights \\ Informing a Human-AI Co‑Debugging System}

\author{Shuyu Shen}
\orcid{0009-0004-7142-6508}
\affiliation{%
  \institution{The Hong Kong University of Science and Technology (Guangzhou)}
   \city{Guangzhou}
  \country{China}
}
\email{sshen190@connect.hkust-gz.edu.cn}

\author{Sirong Lu}
\orcid{0009-0000-3883-1449}
\affiliation{%
  \institution{The Hong Kong University of Science and Technology (Guangzhou)}
   \city{Guangzhou}
  \country{China}}
\email{slu075@connect.hkust-gz.edu.cn}

\author{Leixian Shen}
\authornote{Yuyu Luo and Leixian Shen are corresponding authors.}
\orcid{0000-0003-1084-4912}
\affiliation{%
  \institution{The Hong Kong University of Science and Technology}
  \city{Hong Kong SAR}
  \country{China}
}
\email{lshenaj@connect.ust.hk}

\author{Yuyu Luo}
\authornotemark[1]
\orcid{0000-0001-9530-3327}
% \authornote{Co-corresponding author} 
\affiliation{%
  \institution{The Hong Kong University of Science and Technology (Guangzhou)}
   \city{Guangzhou}
   \country{China}
}
\email{yuyuluo@hkust-gz.edu.cn}

\renewcommand{\shortauthors}{Shen et al.}

\begin{abstract}

Visualization authoring is an iterative process requiring users to adjust parameters to achieve desired aesthetics. Due to its complexity, users often create defective visualizations and struggle to fix them. Many seek help on forums (\eg Stack Overflow), while others turn to AI, yet little is known about the strengths and limitations of these approaches, or how they can be effectively combined. We analyze Vega-Lite debugging cases from Stack Overflow, categorizing question types by askers, evaluating human responses, and assessing AI performance. Guided by these findings, we design a human-AI co-debugging system that combines LLM-generated suggestions with forum knowledge. We evaluated this system in a user study on 36 unresolved problems, comparing it with forum answers and LLM baselines. Our results show that while forum contributors provide accurate but slow solutions and LLMs offer immediate but sometimes misaligned guidance, the hybrid system resolves 86\% of cases, higher than either alone.
\end{abstract}


% \ccsdesc[500]{Human-centered computing~Human computer interaction (HCI)}

\vspace{-5px}
\begin{teaserfigure}
  \includegraphics[width=0.99\textwidth]{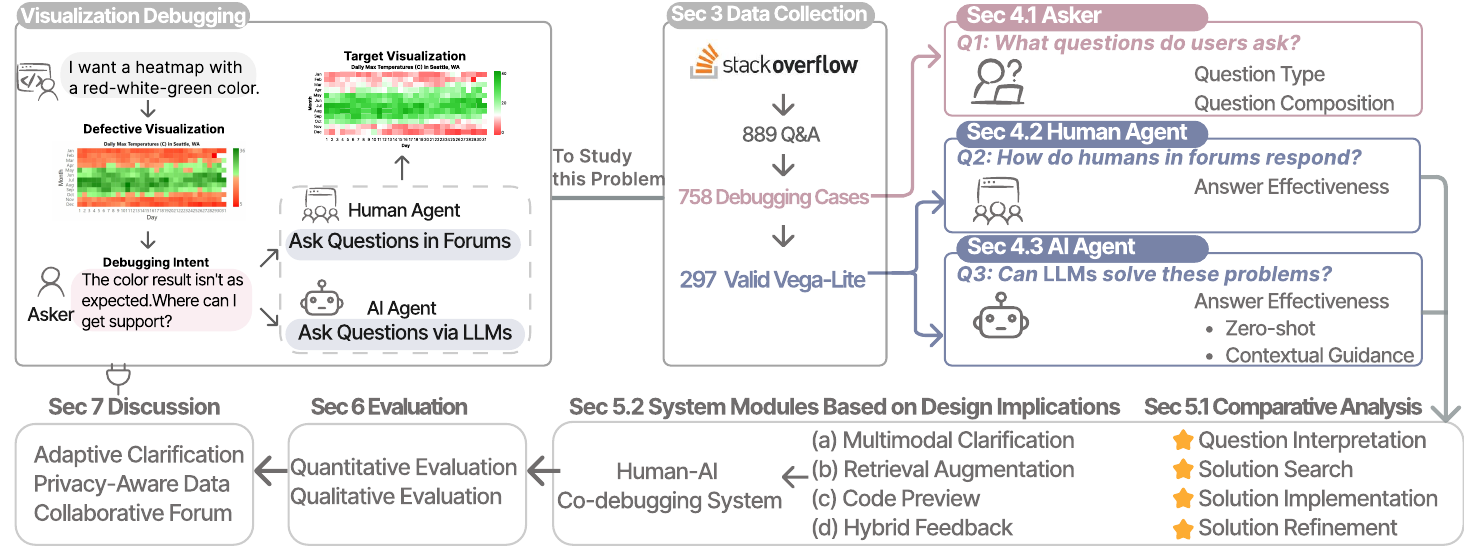}
  \vspace{-5px}
  \caption{The research overview of the paper. To study the problem of visualization debugging (user-identified issues in defective visualizations), we first filter out Vega-Lite debugging Q\&A on Stack Overflow (Section~\ref{sec:dataset}). We then analyze research questions from three perspectives: askers, human agents, and AI agents (Section~\ref{sec:human}). For the human and AI agents, we summarize their comparative findings and use these as design implications for a human-AI co-debugging system (Section~\ref{sec:System}). 
  Finally, we evaluate the system through a user study (Section~\ref{sec:Experimental Evaluation}) and discuss broader implications, limitations, and future opportunities (Section~\ref{sec:discussion}).
  }
  \Description{.}
  \label{fig:teaser}
\end{teaserfigure}


\maketitle

\section{Introduction}
\label{sec:introduction}

Visualization authoring is an iterative and challenging process. Users often need to modify intermediate visualization results to achieve their desired aesthetics and functionality~\cite{DBLP:journals/tvcg/LiLFZLL24, vistalk, NLISurvey, deepeye_icde, DBLP:journals/pacmmod/LuoZ00CS23, DBLP:conf/sigmod/Luo00CLQ21,DBLP:journals/pvldb/LuoLFCT25,DBLP:journals/tkde/LuoQCTLL22,luo2021natural}. This process requires adjusting parameters such as color schemes, scales, and layouts to convey data insights effectively~\cite{DBLP:journals/corr/abs-2406-11033,DBLP:journals/vldb/QinLTL20, DBLP:journals/tkde/LiuSLMJZFLTL25}.
In programming-based authoring tools, such as \vl~\cite{Satyanarayan2017}, users define visualizations through declarative grammar. 
For example, as shown in Figure~\ref{fig:teaser}, a user employs Vega-Lite to create a heatmap but struggles to adjust the color scheme to a specific red-white-green gradient. Despite thoroughly consulting the \vl documentation~\cite{vega_lite_docs}, and interacting with the compiler~\cite{vega_editor}, the user may still not be able to resolve this issue independently. The key challenge is that visualization debugging relies on human visual perception, which is fundamentally different from traditional code debugging~\cite{liu2025nl2sql}.
Given the complexity of these adjustments, users often seek external help with \textit{visualization debugging}~\cite{Satyanarayan2019a}. 

Traditionally, users turn to online Q\&A forums like Stack Overflow for help, which provides personalized, context-aware advice from experienced peers~\cite{Battle2022a}.
However, the quality of responses can vary widely depending on the expertise of the responders, and users may experience delays before receiving effective feedback. In recent years, Large Language Models (LLMs) have emerged as a new avenue for support, offering immediate feedback to correct defective visualizations~\cite{DBLP:journals/pvldb/LiLCLT24, DBLP:journals/corr/abs-2510-23587, DBLP:journals/corr/abs-2510-17586, DBLP:conf/icml/Li0FXC0L25}.
However, LLMs are prone to generating incorrect or irrelevant solutions due to hallucination issues~\cite{DBLP:journals/corr/abs-2410-10762,DBLP:conf/cidr/0001YF0LH24} and often require additional contextual guidance, such as external documentation, to provide reliable answers~\cite{Chen2024a, DBLP:conf/emnlp/WuYSW0L24,DBLP:journals/pvldb/YangLCFCT25}.

While visualization debugging is frequent and challenging, there is still a lack of systematic understanding of the strengths and limitations of human assistance and LLM assistance. Moreover, it remains unclear how to best support users when visualizations fail, whether by leveraging community expertise, AI automation, or a combination of both.

To explore these human-AI interaction challenges in real-world scenarios, we study debugging cases from Stack Overflow, a forum where users often seek help for failed visualizations. Building on these cases, we conduct a systematic empirical study that traces the process from user inputs to responses by human and AI agents, focusing on three research questions:

\textit{\textbf{Q1 (Askers)}: What questions do users commonly ask in the Q\&A forum?}
We categorize common challenges faced by users to better understand the gaps between user-created defective visualizations and their desired outcomes.

\textit{\textbf{Q2 (Human Agents)}: How effective are human responses in forums at addressing users' questions?}
We investigate the strengths and limitations of human-provided answers, focusing on whether the proposed solutions actually resolve users' issues.

\textit{\textbf{Q3 (AI Agents)}: To what extent can LLMs provide comparable or improved debugging support for the same questions?}
We evaluate whether LLMs can match or surpass the quality of human responses in Q\&A forums. 

Building on our empirical findings, we compare human and AI perspectives, propose design implications for human-AI co-debugging, and develop a co-debugging system to combine human interpretive strengths with AI's rapid generative capabilities.

\stitle{Contributions.} Our contributions are summarized as follows:

\begin{itemize}
\item \textbf{Dataset Curation.} 
We curate a dataset of 297 Vega-Lite debugging cases from Stack Overflow, covering diverse visualization debugging tasks (single-view, layered, interactive, and multi-view plots) that go beyond basic usage (Section~\ref{sec:dataset}).
\item \textbf{Empirical Study and Findings.} We conduct a systematic comparison of visualization debugging from three perspectives: askers (question patterns), human agents (answer effectiveness), and AI agents (model performance under various prompt settings), revealing the respective strengths and limitations of each (Section~\ref{sec:human}). For example, we found AI agents introduce unrequested aesthetic modifications that diverge from the asker's intent.
\item \textbf{A Mixed-Initiative Human-AI Co-Debugging System.} Building on our empirical findings, we propose design implications for human-AI co-debugging and develop a mixed-initiative co-debugging system (Section~\ref{sec:System}).
We evaluate the system in a user study on 36 previously unsolved problems, comparing its performance to forum answers and LLM-chat baselines. The hybrid system resolves 86\% of cases, higher than either alone
(Section~\ref{sec:Experimental Evaluation}). 
\end{itemize}

\section{Related Work}
\label{sec:related}
This section reviews prior research on programmers' help-seeking behaviors, visualization authoring, and visualization debugging.

\subsection{Help-Seeking of Programmers}
\stitle{Q\&A Platforms.}
Platforms like Stack Overflow are widely used by programmers to seek help~\cite{vasilescu2013stackoverflow, peterson2019gaze, alshangiti2019developing, Battle2022a}. Battle et al.~\cite{Battle2022a} demonstrated the effectiveness of Q\&A platforms in resolving D3\cite{Bostock2011} visualization issues. Despite their benefits, Q\&A platforms present several challenges. Askers are influenced by unfriendly responses~\cite{kabir2024stack, oliveira2018exchange} and gender bias~\cite{ford2017someone}. Responders face challenges such as a lack of knowledge~\cite{rosen2016mobile} and low self-confidence~\cite{oliveira2018exchange}, which can affect answer quality and quantity.
Recent work by Kabir et al.~\cite{kabir2024stack} compared LLMs and human responses for general coding questions, highlighting the risk of misinformation but without addressing how such issues might be mitigated.
Similarly, Suh et al.~\cite{suh2025human} studied prompting strategies for LLMs but focused on code generation more broadly rather than visualization-specific debugging.
Our work focuses on visualization-specific code and investigates how contextual information can enhance the quality of LLM-assisted debugging, offering actionable insights for human-AI collaboration.

\subsection{Visualization Authoring}
\stitle{Programming-Based Authoring.}  
Visualization authoring languages can generally be categorized into imperative and declarative languages. Imperative languages (e.g., Processing~\cite{Reas2006}, ProtoVis~\cite{Bostock2009}, D3~\cite{Bostock2011}) offer flexibility but tend to be cumbersome to use.
Declarative grammars provide varying levels of expressiveness while bridging the gap between programming concepts and users. Low-level grammar (\eg Vega~\cite{Satyanarayan2016}) enables highly flexible and expressive visualization authoring but requires detailed specifications. High-level grammar (\eg \vl~\cite{Satyanarayan2017}, ggplot2~\cite{Wickham2010}, VizQL~\cite{Stolte2002b}, ZQL~\cite{Siddiqui2016b}, Atlas~\cite{Liu2021c}, and Atom~\cite{Park2018}) simplifies interactive chart construction by abstracting construction details.
Compared with imperative languages, declarative approaches trade fine-grained control for greater conciseness and expressivity. Many declarative grammars are JSON-specified~\cite{Satyanarayan2016, Satyanarayan2017, Park2018, McNutt2022}, making them intuitive.
This study focuses on Vega-Lite~\cite{Satyanarayan2017} due to its mature ecosystem and broad adoption in visualization~\cite{galvis, Chen2021b, PyGWalker, taskvis, NarrativePlayer, Chenc}.

\stitle{LLM-Based Authoring.}  
LLMs enable users to describe visualizations in natural language and automatically generate specifications~\cite{DBLP:conf/emnlp/WuYSW0L24, Author2023, Wang2024e, Bendeck2024, Wang2024d}. Tools such as LIDA~\cite{Dibia2023}, ChartGPT~\cite{Tian2023}, and LLM4Vis~\cite{Wang2023c} demonstrate grammar-agnostic generation capabilities. However, studies show that these models often produce defective visualizations~\cite{Chen2024a, srinivasan2021collecting, Ponochevnyi2024, ko2024natural, DBLP:conf/sigmod/Luo00CLQ21}, especially for tasks beyond simple style edits~\cite{hong2023conversational, han2023chartllamamultimodalllmchart, xu2024exploring}.  
Rather than testing models in controlled literacy tasks, we evaluate them on real-world forum questions. These scenarios expose concrete implementation gaps.

\subsection{Visualization Debugging}
\label{subsec:Visualization Debugging}
\stitle{Visualization Debugging.}  
Visualization debugging refers to resolving user-identified issues in defective visualizations to achieve desired visualizations~\cite{Bostock2011, Battle2022a,xie2025visjudge}.
Prior research has examined visualization defects arising from both human- and AI-generated visualizations.
From the human perspective, Lo et al.~\cite{Lo2024a, Lo2024} analyzed over 1,000 misleading visualizations to develop a taxonomy of issues, while Lan et al.~\cite{Lan2024} categorized 2,227 flawed visualizations into misinformation, uninformativeness, and unsociability.
From the AI perspective, studies have assessed both LLM-generated errors~\cite{Choi, Chen2024a} and models' ability to detect misleading charts~\cite{Bendeck2024, Lo2024}.
Debugging tools such as VizLinter~\cite{Chen2021b}, Bavisitter~\cite{Choi}, MisVisFix~\cite{das2025misvisfix}, MobileVisFixer~\cite{wu2020mobilevisfixer}, and GeoLinter~\cite{lei2023geolinter} have been developed to detect and fix non-compliant charts. However, most approaches assume access to well-specified code and primarily focus on literacy-driven rules.  
Our work differs by examining real-world, human-authored debugging cases in forums and designing a mixed-initiative human-AI co-debugging system that directly supports this practice.

\section{Dataset and Experimental Setup}
\label{sec:dataset}
\begin{figure*}[t!]
	\centering
\includegraphics[width=1\linewidth]{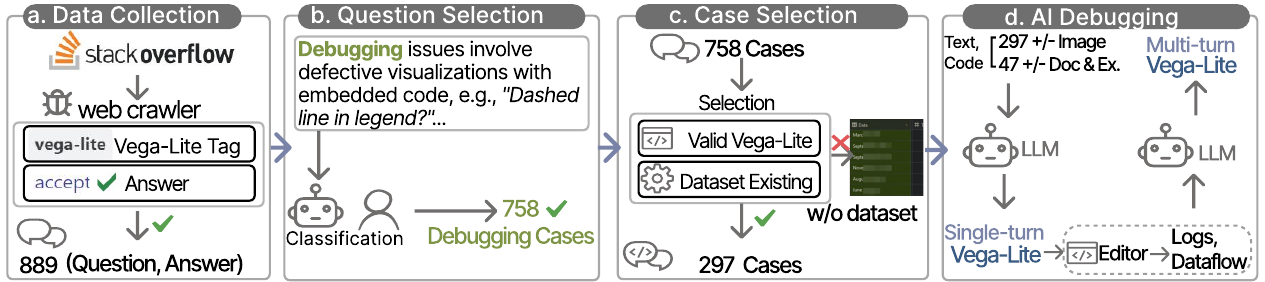}
        \vspace{-1em}
	\caption{Overview of the data curation process, aligned with our research questions (Q1-Q3). It involves four main steps: (a) Data Collection, (b) Question Selection, (c) Case Selection, and (d) AI Debugging.}
	\label{fig:dataset curation}
	% \vspace{-1em}
\end{figure*}

\subsection{Dataset Curation for Visualization Debugging Cases}
\label{subsub:dataset Curation}
This subsection describes the dataset curation process, which consisted of three main steps as illustrated in Figure~\ref{fig:dataset curation}.

\stitle{Data Collection from Q\&A Forums.} Stack Overflow\footnote{\url{https://stackoverflow.com/questions/56425430}.} contains many Q\&A cases related to visualization debugging, covering both coding errors and operational challenges~\cite{Battle2022a}.
These cases often involve multimodal question descriptions, combining text-based issue descriptions, defective visualizations, code snippets, and datasets. As shown in Figure~\ref{fig:dataset curation}(a), we initially crawled 889 cases tagged with \textit{\vl} from Stack Overflow. Each case consists of a question description and its corresponding best answer, which is selected by the question owner.

\stitle{Selection of Debugging Questions.} 
To analyze how users formulate debugging questions, we selected relevant cases from the 889 collected. 
First, we used GPT-4o with a one-shot prompt to perform an initial classification, building on prior research definition~\cite{Battle2022a, Bostock2011, rosen2016mobile, Author2023, Wang2024e}. Detailed prompt templates are available in Appendix~\ref{sub:prompting}. We then conducted a manual review to correct misclassifications, resulting in 758 debugging-related cases.

\stitle{Validation and Case Selection.} We performed a two-stage validation process on the 758 cases to ensure suitability for evaluating answer accuracy in our experiments (Figure~\ref{fig:dataset curation}(c)). The first stage, a Standard JSON Check, validated the Vega-Lite code syntax provided by users~\cite{pezoa2016foundations}. This process filtered out many cases with basic syntax errors, resulting in 413 valid examples. The second stage, a Data Availability Check, identified 350 cases that included a dataset. We manually reviewed these, removing cases with broken URLs or missing data due to privacy concerns, leaving us with 297 usable cases as our primary dataset. 
This final curated dataset formed the basis for the LLM-based experiment.

\subsection{Experimental Design for LLM-Based Debugging}
This subsection describes the experimental design for LLM-based debugging, using 297 curated cases to collect answers under different prompt settings.

\stitle{Models for Evaluation.}
We incorporated a diverse set of models in our evaluation to assess a wide range of model capabilities, including both open-source and closed-source models. We evaluated both Large Language Models (LLMs) (Qwen 2.5-72B~\cite{qwen2025qwen25technicalreport}, DeepSeek-R1~\cite{guo2025deepseek}) and Multimodal LLMs (MLLMs) (Claude-3.5-Sonnet~\cite{anthropic_claude}, GPT-4o-2024-11-20~\cite{hurst2024gpt}, o1-Pro~\cite{OpenAI_o1}). 
We chose these models for their outstanding performance in code generation and image understanding.
Additionally, we evaluated o1-Pro with and without image input to assess the impact of image input. 

\stitle{Prompt Design.} 
We curated prompts to be consistent and fair across all models, ensuring each model received the same information. 
To ensure the models operated on the exact same information available to human community responders, we utilized raw forum descriptions as prompts without manual refinement. This approach allows us to assess the model's performance on natural, often noisy user queries. For text-only models, images were replaced by placeholders to mark the image location, preserving reading fluency. 
We did not generate artificial image captions to avoid introducing bias into the comparison between model and human capabilities.
To verify that different prompt styles do not substantially affect the validity of the generated code, we conducted a preliminary analysis on a subset of 10 cases. We experimented with various prompt variations, including role-playing prompts and CoT prompts. We found that while prompts influenced the output format (e.g., ensuring completeness), the core code remained largely consistent. Detailed prompt templates are available in Appendix~\ref{sub:prompting}.

\begin{figure}[t!]
	\centering
\includegraphics[width=0.9\linewidth]{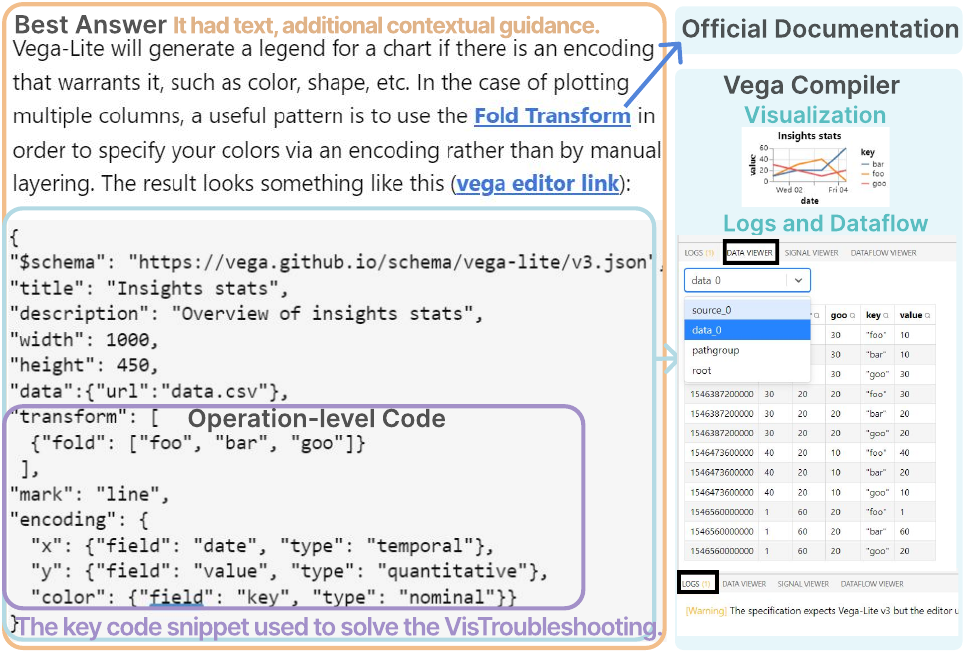}
% \vspace{-1em}
	\caption{An example of a multimodal question related to \vl debugging. \vl specifications can be run in the compiler to obtain the corresponding visualization and compiler feedback.}
	\label{fig:Q2}
	\vspace{-1em}
\end{figure}
\stitle{Experiment Setup.}
To assess LLM effectiveness in debugging and explore the impact of contextual guidance, we designed a Zero-Shot experiment and a series of multi-turn experiments. 
In our Zero-Shot experiment, LLMs using the structured prompt processed 297 cases, generating 1,782 outputs. 
In our multi-turn experiment, inspired by how programmers typically work~\cite{vasilescu2013stackoverflow}, an example is shown in Figure~\ref{fig:Q2}. When programmers view the best answer in a post, they can click the link to navigate to the official documentation. Then, they can run code in the compiler, monitor execution logs, and track intermediate data to debug code. Based on this workflow, we designed two types of contextual guidance: the first type includes logs and dataflow from the compiler~\cite{vega_editor}, and the second type includes documentation and examples~\cite{vega_lite_docs}.
For collection, we crawled the Vega-Lite compiler for logs and dataflow with the Zero-Shot answer. 
Unlike compiler logs and dataflow, relevant documentation was harder to obtain, so we identified suitable references from post discussions and ultimately selected 47 of the 297 cases for our multi-turn experiments.

Details of these 47 cases are provided in Appendix~\ref{sub: 47 cases}. 
Therefore, in our multi-turn experiments, we incorporated two types of additional contextual guidance in the structured prompt: compiler feedback (logs and dataflow) and supplementary resources (documentation and examples). 
We tested six models across various settings, generating a total of 5,346 responses for the experiment.

\subsection{Evaluation Metrics}
\label{sub:evaluation metrics}

\begin{figure}[t!]
	\centering
\includegraphics[width=1\linewidth]{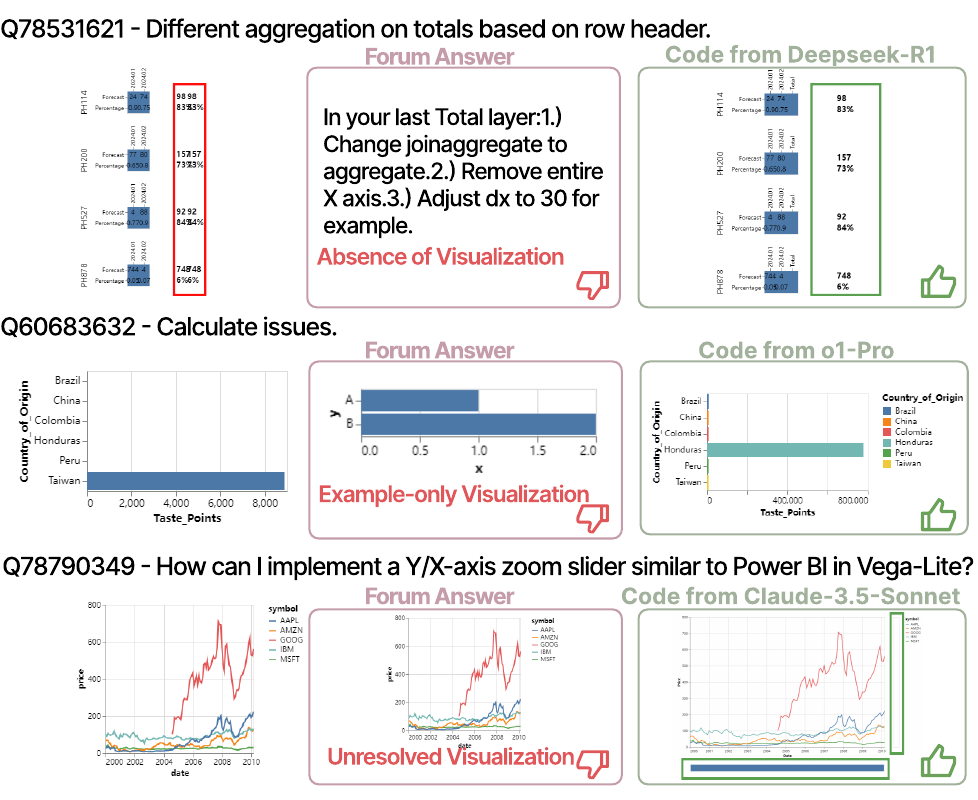}

\caption{Examples of visualization debugging responses across labeling categories.
Red boxes highlight forum answers (Types 1-3) that do not fully resolve the user’s debugging request.
Green boxes show LLM-generated corrections, including correct solutions with minor issues (Type 4, e.g., Q78790349) and fully correct solutions (Type 5, e.g., Q60683632).}
\label{fig:WrongLables}
   \vspace{-1em}
\end{figure}

This subsection presents the evaluation metrics, including labeling categories, annotation procedures, and reliability assessment.

\stitle{Response Labeling Categories.}  
Stack Overflow answers often aim to support community learning by referring to related concepts or reusable examples~\cite{Battle2022a}. Such responses are valuable but may not provide a complete debugging solution. Our evaluation focuses on task-level correctness, whether a response resolves the specific visualization issue with the given data and code.
To align with our visual examples (Figure~\ref{fig:WrongLables}), we define five labeling categories. Types 1-3 refer to responses that do not fully resolve the specific debugging issue and are considered incomplete for the task. Types 4-5 are considered correct, either fully resolving the issue or providing a nearly complete solution with only minor issues.

\utitle{Type 1: Absence of Visualization}. Responses that provide only textual descriptions without rendered visualizations. For example, in Q78531621\footnote{All cases in this paper can be accessed by replacing the question ID at the end of the URL: \url{https://stackoverflow.com/questions/78531621}} (Forum Answer, red box), the solution lists parameter changes in text but provides no visualization, making correctness unverifiable.

\utitle{Type 2: Example-Only Visualization}. Responses that include common examples unrelated to the user's dataset or request. For example, in Q60683632 (Forum Answer, red box), the answer shows a bar chart example that does not adapt to the provided data.

\utitle{Type 3: Unresolved Visualization}. Responses that render a chart but fail to address the core requirement. For example, in Q78790349 (Forum Answer, red box), the user asked for Power BI-style sliders, but the answer only included basic zoom.

\utitle{Type 4: Correct with Minor Issues}. Responses that are largely correct but contain minor flaws. For example, in Q78790349 (Claude-3.5-Sonnet, green box), the generated visualization included a functional slider, but its placement did not follow common standards.

\utitle{Type 5: Correct Visualization}. Responses that fully meet the user's request, render successfully, and align with visualization literacy. For example, in Q60683632 (o1-Pro, green box), the visualization correctly adapted the dataset and resolved the user's issue.

\stitle{Annotation Procedure.}  
Each annotation consisted of the question description and any attached visualizations. Two annotators with 5-6 years of visualization experience independently assessed whether generated code met the requirements, applying the defined labeling categories. Annotators could additionally use LLMs~\cite{OpenAI_o1,anthropic_claude}, official documentation~\cite{vega_lite_docs}, and the compiler~\cite{vega_editor} as references. A custom annotation tool was developed to minimize switching overhead between the IDE and forum~\cite{vasilescu2013stackoverflow,kabir2024stack}.  

\stitle{Assessment of Reliability and Significance.}  
To ensure reliability, we calculated Cohen's Kappa between annotators and consistently obtained scores above 0.80. Disagreements were resolved through discussion. To test the statistical significance of performance differences across models and contextual guidance conditions, we applied Fisher's method to combine p-values, with Bonferroni correction for multiple comparisons.

\section{Empirical Study}
\label{sec:human}
In this section, we explore the roles of askers, human agents and AI agents in visualization debugging through three key aspects: \textbf{(Q1)}~What questions do users ask in the Q\&A forum? \textbf{(Q2)}~How do human agents in forums address debugging issues? and \textbf{(Q3)}~Can LLMs address debugging issues?

\stitle{Interviews with Visualization Practitioners.}
To better understand how real users resolve debugging issues, we conducted semi-structured interviews with six visualization practitioners, comprising three software engineers (I1, I3, I6) and three visualization designers (I2, I4, I5) with 2-5 years of experience using tools (\eg Vega-Lite~\cite{vega_editor}, ggplot2~\cite{Wickham2010}, D3~\cite{Bostock2011} and Tableau~\cite{batt2020learning}). All participants reported regular use of forums (\eg Stack Overflow, GitHub) for debugging and knowledge sharing. Each 40-60 minute interview followed a four-part protocol: (1) participants’ visualization and forum background, (2) a walkthrough of a recent debugging case, (3) how participants locate, evaluate, and combine external resources, and (4) the challenges they face and the support they desire during debugging. Insights from these interviews are integrated into the Q1-Q3 empirical analysis as practitioner perspectives.

\subsection{(Q1) What Questions Do Users Ask in the Q\&A Forum?}
\label{sub:q1}
Building on prior work~\cite{Battle2022a, Bostock2011,kabir2024stack,rosen2016mobile}, we analyze question types and composition using the dataset curated in Section~\ref{sec:dataset}.

\subsubsection{Analysis of Question Types}
\label{subsub: 4.1.1 qtype}
In this section, we analyze the types of questions that askers post on the forum.

\stitle{Debugging issues (758 out of 889 cases)} represent the most common category, indicating a substantial demand for practical visualization solutions. The primary focus lies in enhancing or correcting existing visualizations. This aligns with results by Battle et al.~\cite{Battle2022a}, noting that askers often face unexpected visualizations. 

We further divided debugging issues into four subcategories, following the \vl official documentation~\cite{vega_lite_docs}.

{\underline{View (39.9\%)}} is the most common subcategory, where users ask about modifying visual elements, \eg \textit{``How to get a dashed line in the legend?''}. The user seeks guidance on altering the visual style of chart elements in the legend.  

{\underline{Data Transformation (28.6\%)}} is the second most common subcategory, focusing on questions about data requirements, \eg \textit{``How to encode table-based data?''}. The user seeks guidance to convert a raw table into a format suitable for the desired visualization.

{\underline{Interaction (18.5\%)}} represents the third most common subcategory, covering questions on adding or refining interactive features with dynamic behaviors, \eg \textit{``Is there a way to have a dynamic tooltip?''}. The user has already completed the tooltips but wants them to dynamically display updated values for each line.

{\underline{Composition \& Layout (13.0\%)}} is the least common subcategory, focusing on organizing multiple views and combining visualizations into complex layouts, 
\eg \textit{``How to create scatter matrix?''}. The user has a single-view plot but wants to map the data into a matrix of four charts.

\stitle{Non-Debugging issues (131 out of 889 cases)} primarily include \textit{authoring issues} and \textit{system issues}. In \utitle{authoring issues} (75 out of 889 cases), askers typically provide data and describe requirements, requesting visualizations to be created from scratch, \eg \textit{``How do I create a progress bar?''}. The user is looking for a basic implementation of the bar using embedded data. 
In \utitle{system issues} (56 out of 889 cases), askers often tag their questions with multiple language-related labels that reflect compatibility limitations, \eg \textit{``API misbehaving''}. Such cases typically involve inconsistencies in output during code execution across different environments.

\stitle{Practitioner Perspectives.}
The statistical prevalence of \textit{View} issues aligns with typical development workflows reported by practitioners. Practitioners noted that since complex data transformations are often finalized in upstream tools (\eg SQL or Pandas), their work within visualization tools focuses primarily on visual encoding. Regarding \textit{Interaction} issues, practitioners emphasized that managing dynamic event logic is significantly more time-consuming than creating static plots. Additionally, for \textit{System} issues, they highlighted environmental errors (\eg version incompatibilities) as critical barriers that often force them to abandon the tool.

\subsubsection{Question Composition and External Resources}
\label{subsub:4.1.2}
In this section, we analyzed question composition and the external resources used across 758 debugging issues.

\stitle{Question Composition.} Questions typically combine \textit{code}, \textit{visualizations}, \textit{datasets}, and \textit{textual descriptions} to enhance clarity. The following percentages indicate the proportion of questions containing each component (totals exceed 100\% because questions may include multiple components).

\utitle{(1) Textual descriptions (100\%)} form the primary context, expressing askers' needs, observations, and challenges. 

\utitle{(2) Visualizations (52.5\%)} (\eg png) frequently supplement these descriptions, showing defective, annotated, or desired outputs, which aligns with findings in D3 community~\cite{Battle2022a}.

\utitle{(3) Code (54.5\%)} is frequently included for modification, either embedded directly or linked to the compiler~\cite{vega_editor}. Much of this code is defective, with over 45\% of cases containing syntax errors. 

\utitle{(4) Datasets (39.1\%)} are provided to validate visualizations and guide implementation, appearing as embedded tables or external links. In some cases, the dataset is omitted, leaving responders to infer data structure and content.

\stitle{External Resources.} 28.6\% of askers reference supplementary resources. The primary resources include \textit{documentation, examples}, and \textit{previous Q\&A posts}~\cite{oliveira2018exchange}.

\utitle{(1) Documentation (11.2\%)} includes official documentation and external resources like Altair guides and Wiki pages, helping askers understand syntax and usage. 

\utitle{(2) Examples (6.1\%)} from the official Vega-Lite website serve as customizable templates for implementation across various applications.

\utitle{(3) Q\&A Posts (16.6\%)} include Stack Overflow, GitHub, and other forum posts that provide solutions based on prior solutions shared by the community. 

\stitle{Practitioner Perspectives.}
The observed frequent omission of datasets aligns with practitioner feedback regarding data privacy protocols. Practitioners explained that they are strictly prohibited from sharing proprietary company data. One practitioner (I5) characterized this workflow as blind coding, noting that the lack of real data significantly hinders debugging. This constraint explains the heavy reliance on official examples, as practitioners prefer verified templates that function independently of sensitive internal datasets.

\subsection{(Q2) How Do Human Agents Address Visualization Debugging?}
\label{sub:Q2}

To investigate how human agents address debugging issues, we examine two aspects: the composition of forum responses and the effectiveness of the provided code. 
\subsubsection{Best Answer Composition and Response Time}
\label{subsub:4.2.1}
We analyzed the composition of 758 visualization debugging answers and their temporal characteristics.  

\stitle{Answer Composition.} 
We further examined how best answers are composed, focusing on the inclusion of code, visualizations, and external resources. Similar to question composition, a single answer may contain multiple components, so these percentages do not sum to 100\%. 

\utitle{Complete code (54.5\%)} is the most common, typically including \vl that directly addresses the user's question. Some answers provided links to the compiler (30.9\%), offering an interactive solution.

\utitle{Visualization (60.3\%)} is regularly used in the best answers, providing clarification to the textual description. Visual aids help to demonstrate the output of the code and clarify the debugging process.

\utitle{External resources (36.1\%)} are frequently referenced in the best answers. A significant portion (24.0\%) referenced official or unofficial resources, while 16.0\% cited prior posts. 

\stitle{Response Time and Iterative Process.}
Our analysis reveals that visualization debugging is often an iterative process rather than a single-turn solution. Among 758 cases, 432 required askers to refine their problem statements (supplying missing data or clarifying requirements following responder feedback). Accepted solutions involved an average of 1.77 follow-up comments. The median response time was 3.32 hours for askers and 0.77 hours for responders. This latency, combined with the voluntary nature of forum support, often leads to extended time to obtain a complete solution.

\stitle{Practitioner Perspectives.}
Consistent with the high percentage of complete code in best answers (54.5\%), practitioners emphasized that they need runnable code to verify solutions immediately. Text explanations are not enough. Regarding response time, interviewees corroborated the impact of the observed latency, characterizing it as a bottleneck for urgent tasks. 
Consequently, they prioritize real-time collaboration with colleagues, treating forums as a fallback resource when internal expertise is unavailable.

\subsubsection{The Effectiveness of Forum Answer Code}
\label{subsub:4.2.2 human experiment}
To investigate the effectiveness of forum-provided solutions, we focused on the 297 curated cases introduced in Section~\ref{sec:dataset}, which had passed both syntax and data-availability checks. 
Although these cases correspond to community-recognized best answers, many still contained incomplete specifications or missing data, making manual evaluation challenging. 
For each case, we analyzed code complexity, visualization type, and operational characteristics (Appendix~\ref{sub:297 cases}).

Overall, forum-based solutions achieved an accuracy of 81.4\%. Among the remaining 18.6\% that failed, errors fell into three main categories: \textit{Absence of Visualization} (34.5\% of errors), \textit{Example-Only Visualization} (23.6\%), and \textit{Unresolved Visualization} (41.8\%). Some of these failures were due to time constraints or outdated Vega-Lite versions, which restricted the quality of human-provided solutions.

By operation type, accuracy was lowest for \textit{Pivot} (66.7\%), \textit{Time Unit} (64.7\%), and \textit{Tooltip} (63.2\%). 
By visualization type, accuracy was highest for \textit{Single-View Plots} (82.4\%) and \textit{Composite \& Layered Plots} (83.1\%), but lower for \textit{Interactive Plots} (74.5\%) and \textit{Multi-View Displays} (79.4\%). 
These results suggest that forum contributors face challenges with interactive and multi-view visualizations, which often require managing interactions and cross-view relationships.

\stitle{Practitioner Perspectives.}
The practitioner's emphasis on visual verification aligns with the high rate of ``Absence of Visualization'' errors in our dataset. Practitioners explained that without a rendered image, assessing code correctness becomes inefficient. Furthermore, mirroring the lower accuracy observed in complex visualizations, practitioners noted that adapting intricate code from forums is highly error-prone. Consequently, rather than troubleshooting defective complex code, they often abandon their initial design in favor of adapting a verified example to save time (I2, I4).
\subsection{(Q3) How Do AI Agents Address Visualization Debugging?}
\label{sec:llm}
This section explores how AI agents perform in visualization debugging under various prompt settings.
In our AI agent setting, we simulate an agentic debugging process by integrating the Vega-Lite compiler and retrieval modules into the reasoning loop. As shown in Figure~\ref{fig:dataset curation}(d), we equip the LLM with diagnostic signals (logs, dataflow) and external references (documentation, examples), which are commonly requested during tool-use cycles. This setup allows us to evaluate agentic capabilities by systematically assessing the specific contribution of each information source to the debugging performance.

\subsubsection{Effectiveness of LLMs in Visualization Debugging}
\label{subsub:Q3}
In this section, we evaluated the fundamental capabilities of LLMs in visualization debugging through a Zero-Shot experiment. We found that LLMs demonstrated limited effectiveness, yet they occasionally generated novel solutions that surpassed forum-provided answers.

\begin{figure}[t!]
    \centering
    \includegraphics[width=\linewidth]{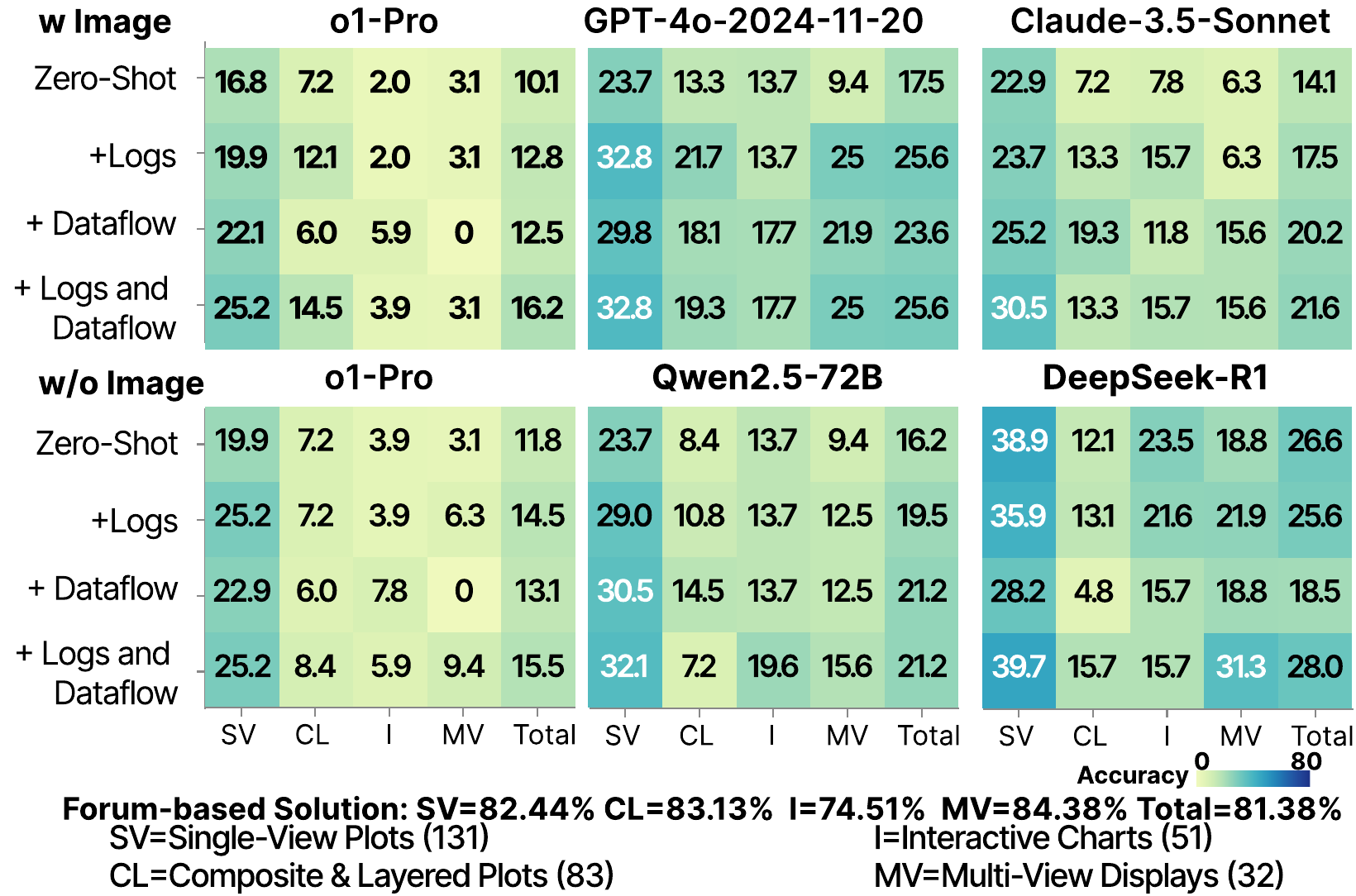}
    \caption{A heatmap visualizing the solution accuracy across 297 cases, comparing four configurations (Zero-Shot, +Logs, +Dataflow, and +Logs + Dataflow) across four visualization types (Q3).}
    \label{fig:Log_dataflow_heatmap}
    \vspace{-0.8em}
\end{figure}

Figure~\ref{fig:Log_dataflow_heatmap} shows Zero-Shot accuracy across 297 cases. 
Within the text-only models, DeepSeek-R1 achieves the highest accuracy (26.6\%). 
Within the multimodal models, GPT-4o-2024-11-20 achieves the highest accuracy (17.5\%).
This indicates that current LLMs, when provided with no additional context, still struggle with the complexity of real-world debugging tasks, highlighting a significant performance gap compared to humans (81.4\%).
Prior research indicates that over half of LLM-generated programming answers contain errors~\cite{kabir2024stack}. Our findings further reveal the limitations of LLMs in visualization debugging.

\stitle{Superior Answers Generated by LLM.}
Within the text-only models, Qwen2.5-72B produced the highest proportion of superior answers (+5.0\%). Within the multimodal models, GPT-4o-2024-11-20 (+4.3\%) and Claude-3.5-Sonnet (+3.7\%) also generated additional correct responses beyond forum solutions.
Three cases are illustrated in Figure~\ref{fig:WrongLables}, where the models successfully modified the code using user-provided data, while the forum failed due to the absence of visualization, example-only visualization, or unresolved visualization.

% \begin{figure*}[t!]
%     \centering
%     \begin{minipage}{0.49\linewidth}
%         \centering
%         \includegraphics[width=1\linewidth]{figures/Log_dataflow_exp_heatmap.pdf}
%         \caption{A heatmap visualizing the solution accuracy across 297 cases, comparing four configurations (Zero-Shot, +Logs, +Dataflow, and +Logs + Dataflow) across four visualization types (Q3).}
%         \label{fig:Log_dataflow_heatmap}
%     \end{minipage}
%     \hfill
%     \begin{minipage}{0.49\linewidth}
%         \centering
%         \includegraphics[width=1\linewidth]{figures/category_heatmap.pdf}
%         \caption{A heatmap visualizing the solution accuracy across 47 cases, comparing seven configurations (Zero-Shot, +Doc, +Ex., +Doc + Ex.,+ Logs,+ Dataflow, and +Doc + Ex. + Logs + Dataflow) across four visualization types (Q3).}
%         \label{fig:category_heatmap}
%     \end{minipage}
%     \vspace{-1em}
% \end{figure*}

\begin{figure}[t!]
    \centering
    \includegraphics[width=\linewidth]{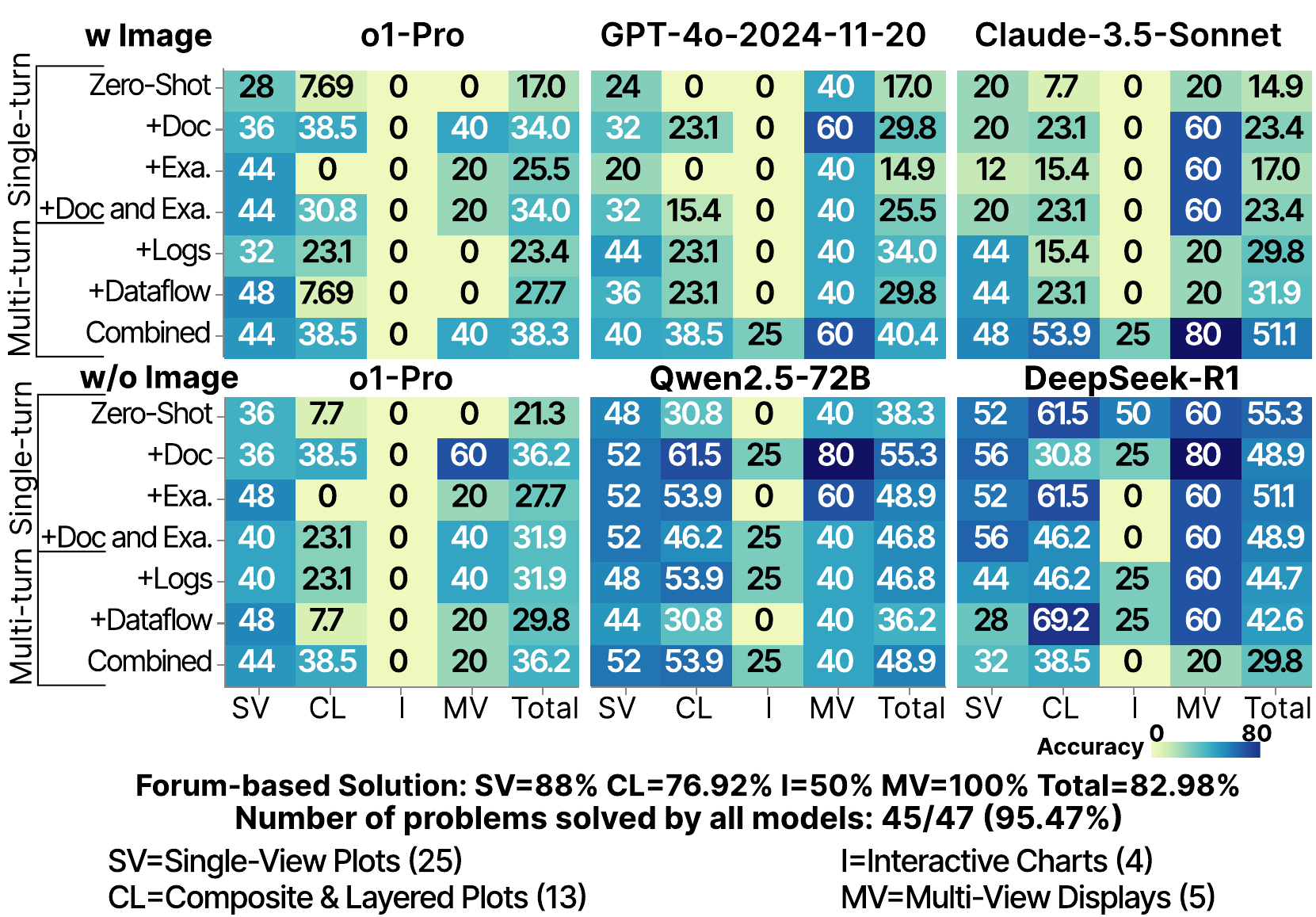}
    \caption{A heatmap visualizing the solution accuracy across 47 cases, comparing seven configurations (Zero-Shot, +Doc, +Ex., +Doc + Ex., +Logs, +Dataflow, and +Doc + Ex. + Logs + Dataflow) across four visualization types (Q3).}
    \label{fig:category_heatmap}
    \vspace{-0.8em}
\end{figure}

\stitle{Practitioner Perspectives.}
Despite the performance gaps observed in our quantitative analysis, some practitioners identified LLMs as their first-pass debugging tool. 
In practice, LLMs help generate initial drafts or explain concepts, but practitioners noted that these drafts often contain version-inconsistent APIs or deprecated syntax, making human verification and logical refinement essential.

\subsubsection{Impact of Additional Contextual Guidance}
\label{subsub: 5.2.1 additional information}

In this section, we analyze the impact of images, compiler feedback and supplementary resources. Figure~\ref{fig:Log_dataflow_heatmap} (the accuracy across 297 cases) and Figure~\ref{fig:category_heatmap} (the accuracy across 47 cases) show the overall impact of additional contextual guidance on LLMs, highlighting that compiler feedback significantly improves accuracy. 

As shown in Figure~\ref{fig:Log_dataflow_heatmap}, all models perform significantly worse than humans (81.4\%) in the four types of visualization debugging cases. The best model, DeepSeek-R1, only achieved 28.0\%. This indicates that current models still struggle to understand and correct visualization errors. LLMs perform best on Single-View Plots (avg 30.9\%) but poorly on Composite \& Layered Plots (avg 13.1\%), Interactive Plots (avg 13.1\%), and Multi-View Displays (avg 16.7\%). These results are far from the performance of forum answers, which achieve significantly higher accuracy in Single-View Plots (82.4\%), Composite \& Layered Plots (83.1\%), Interactive Plots (74.5\%), and Multi-View Displays (84.4\%). 

As shown in Figure \ref{fig:category_heatmap}, integrating additional contextual guidance (documentation, examples, logs, and dataflow) improves models' visualization debugging performance across different model groups. Within the text-only models, Qwen2.5-72B achieved the highest accuracy of 55.3\% with documentation. Within the multimodal models, o1-Pro (38.3\%), GPT-4o-2024-11-20 (40.4\%), and Claude-3.5-Sonnet (51.1\%) reached the highest accuracy with combined information. Considering all experimental results, the highest accuracy reached 95.5\% (45/47), surpassing the forum benchmark of 83.0\%. This demonstrates that LLMs can perform better by offering multiple solutions in complex debugging cases.

\stitle{Impact of Images.} 
For the 297-case dataset, o1-Pro showed slightly higher accuracy when images were not included in the prompt. Performance increased by 1.7\% in the Zero-Shot setting, 1.7\% with Logs, and 0.6\% with Dataflow, while the combined Logs and Dataflow setting showed a small decrease of 0.7\%.
For the 47-case dataset, the model also performed better without images, showing a 2.1\% improvement in the Doc and Example condition.
Overall, these results suggest that including forum images in the prompt did not provide measurable improvement for o1-Pro.

 % In multi-view displays, adding images could slightly hinder the model’s performance.

\begin{figure}[t]
	\centering
\includegraphics[width=1\linewidth]{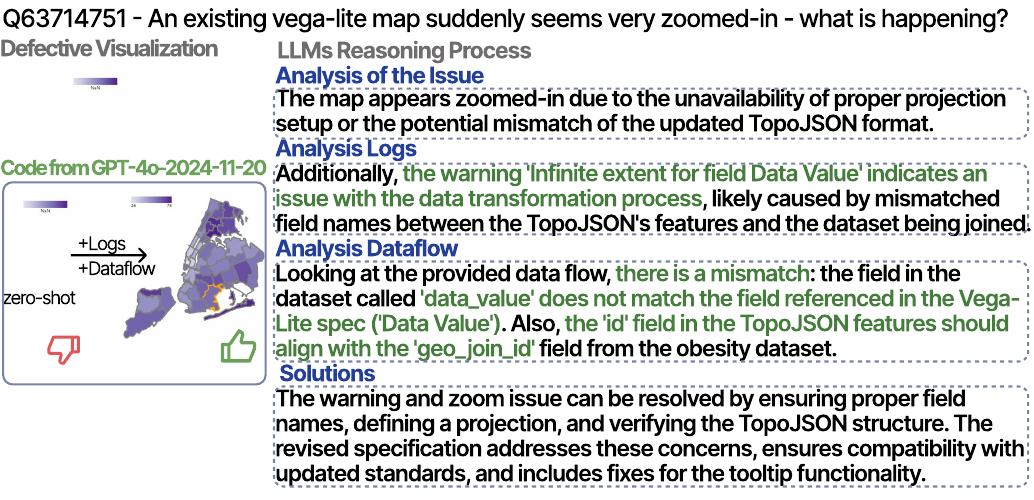}
\vspace{-1em}
   \caption{An example illustrating the impact of additional contextual guidance on LLM responses, with textual reasoning and code visualization (Q3).
}
\label{fig:sample_Logs_dataflow}    
\vspace{-1em}
\end{figure}

\stitle{Impact of Compiler Feedback.} 
Across 297 cases, compiler feedback significantly improved LLMs' visualization debugging performance. Compared to the Zero-Shot baseline (avg 16.1\%), the combination of logs and dataflow achieved the highest and statistically significant improvement (avg 21.4\%), whereas logs alone or dataflow alone produced only non-significant increases. These results suggest that while LLMs frequently capture the problem correctly, logical flaws in the code lead to rendering errors.
A representative example is Q63714751 (Figure~\ref{fig:sample_Logs_dataflow}), where GPT-4o-2024-11-20 initially failed in Zero-Shot. With logs and dataflow, the model identified a warning in the data transformation and detected a data mismatch. This allowed it to ultimately correct the code.

\stitle{Impact of Supplementary Resources.}
Across 47 cases, supplementary resources enhance model performance in visualization debugging. Compared to the Zero-Shot baseline (avg 27.3\%), documentation provided the largest significant improvement (avg 37.9\%). The use of examples alone also showed a positive trend (avg 30.9\%), but this difference was not statistically significant.
Additionally, adding documentation particularly improves performance in Multi-View Displays and Single-View Plots.

\stitle{Practitioner Perspectives.}
Practitioners confirmed the value of compiler feedback, which matches our finding that logs and dataflow improve model accuracy. They explained that as debugging complexity increases, simple code inspection is insufficient. They need intermediate data inspectors and error logs to identify root causes. Regarding supplementary resources, practitioners agreed that official examples are the most critical reference. They typically start with these examples and then leverage LLMs to adapt the code to their specific datasets.

\subsubsection{Characteristics of LLM-assisted Visualization Debugging.}
\label{subsub:Characteristics}
To better understand LLM capabilities in visualization debugging, we analyzed the results through code complexity, operation-level code, and the characteristics of successful cases.

\stitle{Operation-Level Code.}
We evaluated LLM performance by calculating the average accuracy for each operation category. Visualization operations had the highest accuracy (19.7\%), while Composition Layout operations were the lowest (14.4\%). Repeat (33.3\%) and Legend (28.0\%) performed best, while Pivot (6.9\%), Stack (6.0\%), and Impute (0.0\%) were the worst. The model excelled with static visual elements but struggled with complex data transformations (\eg Window, Stack) and layout operations (\eg Facet). This suggests LLMs face challenges beyond just the complexity of the task, likely related to how they use specific visualization commands.

\begin{figure}[t]
	\centering
\includegraphics[width=1\linewidth]{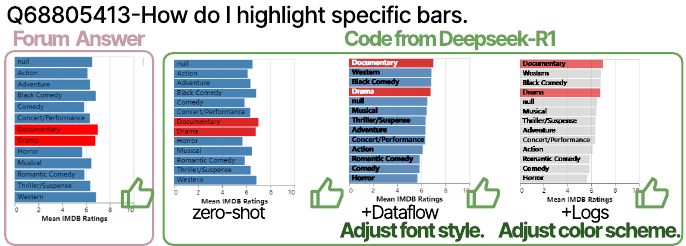}
  \vspace{-1em}

   \caption{The impact of additional contextual guidance on LLM aesthetics compared with forum responses (Q3).}
\label{fig:aesthetic_Sample}
  \vspace{-1em}
\end{figure}

\stitle{Characteristics of Successful Cases.} 
We summarized representative cases where LLMs produced correct solutions. In our analysis of correct labels in 47 cases, we identified 10 instances demonstrating aesthetic modifications.
In Q68805413 (Figure~\ref{fig:aesthetic_Sample}), the user wanted to highlight two specific bars. When dataflow was added, DeepSeek-R1 not only highlighted the bars but also changed the highlighted text to white and bolded all text, enhancing readability. When logs were added, DeepSeek-R1 further adjusted the visualization by graying out the non-highlighted bars, making the red highlights more visible.
This suggests that models can enhance aesthetics, but better control is needed to prevent deviation from the user's intent. 

% For completeness, analyses of unsuccessful cases are presented in Appendix~\ref{sub: 47 cases}.

\stitle{Practitioner Perspectives.}
Aligning with our observation of unrequested aesthetic changes, practitioners noted that AI struggles with precise visual refinement. They explained that while AI easily generates initial charts, fine-grained adjustments (\eg modifying specific legend placements) remain difficult and labor-intensive. To solve this, rather than receiving a single deterministic code snippet which might deviate from their intent, practitioners suggested that models should propose multiple aesthetic alternatives to minimize trial-and-error.

\subsection{Key Findings}
This section summarizes the key findings from our empirical study, which address our research questions on the roles of askers, human agents, and AI agents in visualization debugging.

\stitle{\underline{Askers in the Forum.}} 

We analyzed 889 questions to identify the main challenges faced by askers.
\begin{tcolorbox}
\textbf{Finding 1.1:} Visualization debugging questions from the askers are primarily focused on View and Data Transformation (Section~\ref{subsub: 4.1.1 qtype}).
\end{tcolorbox}

We analyzed 758 debugging cases, focusing on the question components.
\begin{tcolorbox}
\textbf{Finding 1.2:} Questions are typically multimodal, combining textual descriptions and visualizations (\eg png), while the accompanying code is frequently defective and the datasets often incomplete (Section~\ref{subsub:4.1.2}).
\end{tcolorbox}

\stitle{\underline{Human Agents in the Forum.}} 

We analyzed 758 accepted answers to understand how human agents construct their responses.  
\begin{tcolorbox}
\textbf{Finding 2.1:} The composition of human-provided solutions is highly diverse, covering complete code, visualizations, and external resources (Section~\ref{subsub:4.2.1}).
\end{tcolorbox}

We analyzed response times and comments to assess the effort required for producing human responses.
\begin{tcolorbox}
\textbf{Finding 2.2:} Human agents provide high-accuracy solutions but require a high time cost. They also require iterative interaction with askers to clarify problems and refine answers (Section~\ref{subsub:4.2.1}).
\end{tcolorbox}

We evaluated 297 curated cases to examine the effectiveness of answers across different visualization types.
\begin{tcolorbox}
\textbf{Finding 2.3:} Human agents often provide suboptimal guidance for complex visualizations, such as interactive Charts and Multi-View Displays (Section~\ref{subsub:4.2.2 human experiment}).
\end{tcolorbox}

\stitle{\underline{AI Agents in Visualization Debugging.}}

We evaluated Zero-Shot experiments on 297 cases to test LLMs without contextual guidance.
\begin{tcolorbox}
\textbf{Finding 3.1:} AI agents underperform humans in Zero-Shot visualization debugging, but have the potential to generate novel solutions beyond human-provided answers (Section~\ref{subsub:Q3}).
\end{tcolorbox}

We compared models with and without images. 
\begin{tcolorbox}
\textbf{Finding 3.2:} Although AI agents support long-context reasoning, incorporating forum images does not appear to improve performance and may introduce minor noise (Section~\ref{subsub: 5.2.1 additional information}).
\end{tcolorbox}

We compared models with and without compiler feedback, as well as with and without supplementary resources.
\begin{tcolorbox}
\textbf{Finding 3.3:} AI agents show remarkable performance improvement when augmented with compiler feedback and external documentation (Section~\ref{subsub: 5.2.1 additional information}).
\end{tcolorbox}

We compared operation-level accuracy to identify operations with consistently low accuracy. 
\begin{tcolorbox}
\textbf{Finding 3.4:} AI agents face the challenge of interpreting composition layout and handling specific visualization commands, such as Pivot, Stack, and Impute (Section~\ref{subsub:Characteristics}).
\end{tcolorbox}

We analyzed cases where LLMs introduced modifications not requested by askers.
\begin{tcolorbox}
\textbf{Finding 3.5:} AI-generated solutions sometimes introduce unrequested aesthetic adjustments. Better control is needed to prevent deviation from the askers' intent (Section~\ref{subsub:Characteristics}).
\end{tcolorbox}

Overall, these findings demonstrate the complementary trade-offs between humans and AI. Humans are proficient in interpretive accuracy but are slow, while AI offers scalability and speed but requires contextual guidance and validation. 
Given these challenges, particularly the frequent unintended aesthetic changes (Finding 3.5) and the difficulty of expressing complex intent through textual instructions (Finding 1.2), fully automated agentic pipelines remain unreliable for visualization debugging in practice. This underscores the need for a human-in-the-loop approach where the system guides the prompting process, ensuring reliability despite variability in human input quality.
This complementarity motivates the design of our mixed-initiative system.

\section{Mixed-Initiative Human-AI Co-Debugging System}
\label{sec:System}
To translate our empirical findings into the design of our mixed-initiative system, we employed a deductive coding protocol to derive Design Implications through three steps.
(1) By integrating the search-and-refine workflow from visualization study~\cite{Battle2022a} with the intent-and-generation cycle from AI programming study~\cite{kabir2024stack}, we established four distinct debugging stages: \textit{Question Interpretation}, \textit{Solution Search}, \textit{Solution Implementation}, and \textit{Solution Refinement}.
% Specifically, \textit{Question Interpretation} and \textit{Solution Implementation} align with Kabir et al.'s focus on intent understanding and generative errors, while \textit{Solution Search} and \textit{Solution Refinement} draw from Battle et al.'s findings on example retrieval and iterative visual debugging.
(2) We mapped each empirical finding (Findings 1.1-3.5) and practitioner perspective to the relevant stages.
(3) We synthesized these findings to compare the strengths and limitations of human and AI agents at each stage. This comparison revealed specific capability gaps in current workflows, which directly motivated the four Design Implications detailed below and summarized in Table~\ref{tab:capabilities_comparison}.

\stitle{Question Interpretation.}
Human agents are effective in intent disambiguation through iterative dialogue (Finding 2.2). However, the prevalence of defective code and incomplete datasets forces a manual reconstruction of the problem context (Finding 1.2).
AI agents can rapidly parse long-context descriptions but struggle to extract precise constraints from noisy multimodal inputs (Finding 3.2).
Therefore, the design must bridge AI's processing speed with human verification to reduce initial ambiguity.

\utitle{Design Implication 1 (DI1):} The system should parse multimodal input into unified structured representations, enabling users to verify and clarify intent before solution generation.

\stitle{Solution Search.}
Human agents ensure quality by relying on grounded external resources (Finding 2.1), but utilizing these sources involves a manual, sequential workflow that is inefficient (Finding 2.2).
Conversely, AI agents demonstrate the potential to synthesize novel solutions (Finding 3.1), but may produce version-inconsistent APIs when lacking proper context (Practitioner Perspectives from Section~\ref{subsub:Q3}).
Therefore, the design must leverage AI to accelerate discovery while strictly grounding its outputs in verified external resources to ensure correctness.

\utitle{Design Implication 2 (DI2):} The system should incorporate diversified retrieval augmentation that leverages AI's broad synthesis capacity while grounding outputs in relevant human knowledge.

\stitle{Solution Implementation.}
While human agents provide contextually correct logic, they often supply incomplete snippets or links rather than full executable code, shifting the implementation burden to the asker (Finding 2.3).
AI agents can instantly generate complete code iterations (Finding 3.3) but are prone to syntax errors in complex composition and data transformation (Finding 3.4).
Therefore, the design must leverage AI to bridge the gap between abstract snippets and executable code, while providing instant feedback for error detection.

\utitle{Design Implication 3 (DI3):} The system should facilitate instant code generation with real-time preview, enabling efficient implementation while allowing users to validate solutions visually.

\stitle{Solution Refinement.}
Human experts rely on visual inspection to verify correctness (Practitioner Perspectives from Section~\ref{subsub:4.2.2 human experiment}). However, articulating aesthetic preferences or fine-grained adjustments via text is imprecise and labor-intensive, leading to inefficient refinement loops (Finding 2.2).
AI agents can efficiently debug logic when guided by compiler feedback (Finding 3.3), but often introduce unrequested aesthetic changes (aesthetic drift) that deviate from user intent (Finding 3.5).
Therefore, the design should overcome the ambiguity of text descriptions by enabling direct visual guidance.

\utitle{Design Implication 4 (DI4):} The system should establish a feedback loop that combines AI-driven debugging with human \textit{visual-spatial} input to minimize textual ambiguity.

\begin{table*}[t!]
\caption{Comparison of human and AI agents across four Q\&A stages in visualization debugging, with corresponding design implications (a-d). }
\label{tab:capabilities_comparison}
\centering
\renewcommand{\arraystretch}{1.3}
\begin{tabular}{m{2.2cm}m{4cm}m{4cm}m{5cm}}
\toprule
\textbf{Q\&A Stage} & \textbf{Human Agent} & \textbf{AI Agent} & \textbf{Design Implication} \\
\midrule

\makecell{Question\\ Interpretation} & 
Effective at clarifying intent through iterative dialogue (Finding 2.2). However, manual reconstruction of context from defective code and missing data is inefficient (Finding 1.2). &
Can leverage long context to infer needs from vague descriptions. However, struggles to extract precise constraints from multimodal noise (Finding 3.2). &
\textbf{a. Multimodal Clarification:} Parse multimodal input into structured intent for user verification before solution generation.\\
\midrule 

\makecell{Solution\\ Search} & 
Ensure quality by retrieving grounded external resources (Finding 2.1). However, this manual, sequential search is inefficient (Finding 2.2). &
Demonstrate potential for novel solutions (Finding 3.1). Can rapidly synthesize broad sources, though prone to hallucinations without grounding (Practitioner Interviews). &
\textbf{b. Retrieval Augmentation:} Leverage AI’s ability to integrate information broadly while grounding solutions in validated community knowledge.\\
\midrule 

\makecell{Solution\\ Implementation} & 
Provide conceptually correct logic, but may offer incomplete snippets or links rather than full executable code (Finding 2.3). &
Instantly generate complete code iterations (Finding 3.3), yet prone to errors in composition layout and specific commands (Finding 3.4). &
\textbf{c. Code Preview:} Provide real-time code generation and preview for rapid visual validation.\\
\midrule 

\makecell{Solution\\ Refinement} & 
Effective at visual inspection to verify correctness (Practitioner Interviews). However, articulating spatial adjustments via text is imprecise and inefficient (Finding 2.2). &
Can improve logic using compiler feedback (Finding 3.3), yet tendency to introduce unrequested aesthetic changes deviates from intent (Finding 3.5). &
\textbf{d. Hybrid Feedback:} Establish a feedback loop that enables visual-spatial input to constrain AI generation.\\
\bottomrule
\end{tabular}
 \vspace{-1em}
\end{table*}

\begin{figure*}[t!]
	\centering
\includegraphics[width=1\linewidth]{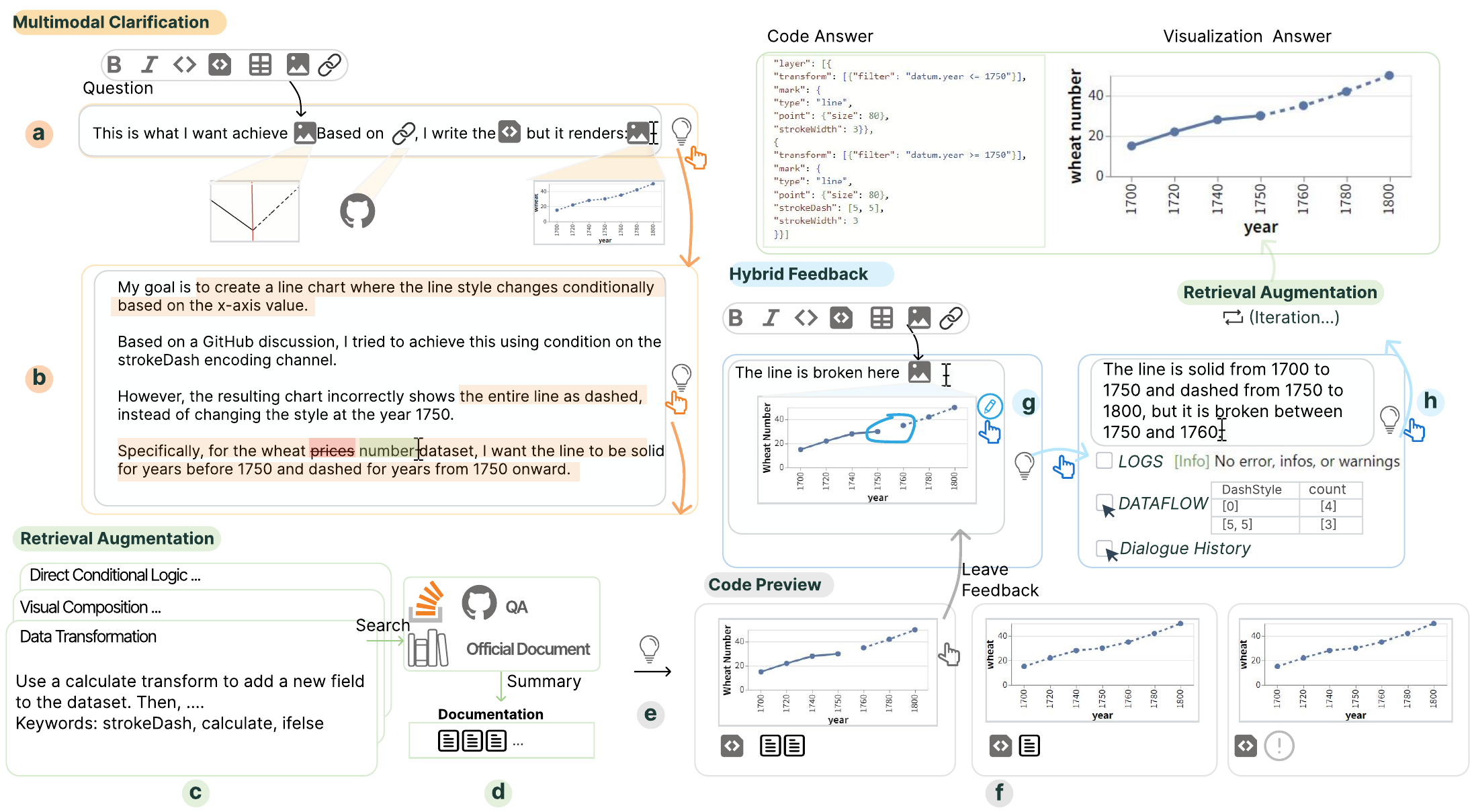}
  \vspace{-1em}
  % \vspace{-2em}
   \caption{An overview of the mixed-initiative human-AI co-debugging system workflow. The process begins with Multimodal Clarification, where the system guides users to refine their problem description. Next, Retrieval Augmentation searches for relevant information from external knowledge sources, which is then used by Code Preview to produce and present diverse solutions. Finally, Hybrid Feedback enables users to provide refinements through textual-visual annotations, closing the interactive debugging loop.}
\label{fig:Demo}
    \vspace{-1em}
\end{figure*}

\subsection{System Modules Based on Design Implications}
Our system operationalizes the four design considerations into an integrated workflow (Figure~\ref{fig:Demo}(a-b)) within a conversational interface to support mixed-initiative human-AI collaboration.

\subsubsection{Multimodal Clarification (Addressing DI1).}
To disambiguate user intent before solution generation, our system implements a Multimodal Clarification function. 

Users can provide multimodal input, including problem descriptions, defective code, and images, through an interface that supports inline image embedding (a). An MLLM parses this input by converting visual information into natural language descriptions and integrating them into a unified textual representation (b). To better leverage images without being affected by potential noise (Finding 3.2), the model highlights its image-derived inferences within the text (e.g., tagged spans), enabling users to verify and correct them. When inconsistencies are detected, users can iteratively refine the description until the intent is accurately captured. 
This mechanism preserves the human advantage of precise intent understanding through iterative dialogue (Finding 1.2), while mitigating the limitations of human-only interaction, such as high time cost (Finding 2.2). Ultimately, it establishes a shared understanding between the user and AI agent, thereby reducing ambiguity and enhancing the reliability of subsequent debugging steps.

\subsubsection{Retrieval Augmentation (Addressing DI2).}
To overcome the limitations of sequential human search, our system implements the Retrieval Augmentation function.

The process begins by having an LLM generate three distinct exploratory plans based on the clarified user intent (c), leveraging AI's capacity for broad knowledge synthesis and novel solution generation (Finding 3.1). Each plan is then used to retrieve relevant documents from a knowledge base comprising Vega-Lite official documentation, GitHub issues, and Stack Overflow posts, thereby capturing the diversity and quality of human-provided solutions (Finding 2.1) while avoiding the efficiency limitations of sequential search (Finding 2.2). For each plan, the five most relevant documents are retrieved using an embedding-based similarity search over the vector database.
Finally, the LLM synthesizes the retrieved documents, the original plan, and the user's intent to generate a concise technical summary for each proposed solution (d), thus extending the diversity of human-provided solutions with AI-driven knowledge integration.

\subsubsection{Code Preview (Addressing DI3).}
To combine the speed of AI-generated code with the reliability of implementation validation, our system implements the Code Preview function.

Building on Finding 3.3, which shows that external knowledge improves accuracy, the system leverages retrieved and synthesized resources from the retrieval augmentation process (e), together with the clarified user intent, to prompt the LLM. The model then produces executable code, directly addressing the latency in human refinement processes (Finding 2.2) and the difficulty of handling complex visualizations (Finding 2.3). To mitigate AI's difficulty in interpreting composition layouts and specific visualization commands (Finding 3.4), the system incorporates real-time rendering, allowing users to immediately validate each proposed solution (f). If satisfied, users can copy the generated code for use, with relevant documentation links provided for further exploration.

\subsubsection{Hybrid Feedback (Addressing DI4).}
To enable an efficient refinement loop that combines human perceptual feedback with AI-driven debugging, our system implements Hybrid Feedback.

When users wish to further explore a particular solution from code preview, they can click on the visualization to access an embedded Vega-Lite editor (g). Within this editor, users can debug the code and view real-time logs and dataflow visualizations. If users prefer AI-assisted modifications, they can provide multi-modal input by sketching on the visualization and editing the accompanying text. In addition, compiler logs and dataflow can be included as supplementary signals for the LLM. This design aligns with Finding 3.3, which shows that AI agents improve substantially when augmented with compiler feedback.

The system then integrates the annotated image, modified text, current Vega-Lite specification, and the user's original question intent (from multimodal clarification). Based on this combined input, the LLM generates a feedback intent. Users can edit this generated intent before finalizing it (h). This process retains the human strength of clarifying problems through iterative interaction (Finding 2.2) while mitigating the risk of AI-generated unauthorized modifications (Finding 3.5).

The finalized feedback intent, together with the original question intent and the current Vega-Lite specification, is consolidated into a structured historical context. This context is then passed into retrieval augmentation, supporting subsequent rounds of debugging.

\section{Evaluation}
\label{sec:Experimental Evaluation}
We conducted a user study to evaluate whether the human-AI co-debugging system improves visualization debugging in practice.

\subsection{User Study}
\stitle{Participants.} 
The study recruited 12 participants (6 females, 6 males, aged 20-30), comprising 10 with computer science backgrounds (7 PhD, 2 Master's, 1 undergraduate) and 2 from applied domains. All participants received a \$20 gift card as compensation. 

In terms of expertise, 6 were visualization or HCI researchers who frequently debug visualizations weekly or more often (P1, P2, P5, P8, P9, P10). 2 were AI researchers who engaged in algorithm debugging almost daily as part of their research (P11, P12). 2 were software engineers with regular experience in software debugging on a weekly basis (P3, P6). 2 were product managers who only occasionally performed visualization debugging, typically monthly or less (P4, P7).  
On average, participants had approximately 4-5 years of experience in code debugging, though the depth of practice varied across domains.

\stitle{Dataset.}
We used a dataset of 36 Vega-Lite debugging questions from Stack Overflow. Each case contained at least one defective visualization. All selected questions had no accepted answer at the time of collection (see Appendix~\ref{sub:Test Dataset}), ensuring that the tasks represented realistic debugging challenges.
From an initial collection of 177 unresolved posts, we randomly sampled 36 cases that satisfied two criteria: (1) the post addresses a visualization debugging problem, and (2) it provides a standard Vega-Lite JSON specification with accessible data sources (Section~\ref{subsub:dataset Curation}).
The cases covered a representative range of difficulties, from simple edits (\eg changing bar colors) to complex modifications (\eg debugging multi-facet layouts or resolving overlapping encodings).

\stitle{Tasks and Procedure.}
Before the experiment, participants completed a 10-minute tutorial on Vega-Lite syntax and a 10-minute demonstration of the system, with opportunities to ask questions and practice simple tasks.
In the main experiment, we adopted a within-subjects design with counterbalancing (Latin square). Each participant was randomly assigned 6 cases from the dataset: three under the LLM Interaction condition and three under the Human-AI Co-Debugging System condition. 
The two conditions were defined as follows: 
(i) \textit{LLM Interaction}: participants interacted with GPT-4o using a standard chat-based interface; 
(ii) \textit{Human-AI Co-Debugging System}: participants used our system with multimodal clarification, retrieval augmentation, code preview, and hybrid feedback. 
In the LLM Interaction condition, participants interacted with GPT-4o via a standard chat interface. 
To ensure a fair comparison, the initial input prompt was identical across both conditions, after which participants engaged in iterative prompting based on the model's feedback.
To maintain a realistic debugging loop, they were provided with the official Vega-Lite online editor for real-time rendering and log inspection, and were explicitly permitted to paste code snippets and compiler error messages back into the chat. They also had access to a web browser for searching official documentation and forum posts. 
This setup reflects a common real-world workflow observed in our practitioner interviews (Section~\ref{sec:human}), in which developers combine an LLM assistant, an executable visualization editor, and web resources when debugging. 
For each case, participants had up to 5 minutes to read and understand the problem, followed by up to 5 minutes to solve it under the assigned condition.

\stitle{Evaluation.} Three visualization researchers independently reviewed all submitted solutions. 
Raters were blinded to condition, and evaluations were based on the rendered charts and Vega-Lite specifications. 
Correctness was defined as whether the final visualization matched the problem requirements. 
Inter-rater reliability before discussions was high (Krippendorff’s $\alpha$ = 0.82). 
Disagreements were resolved through discussion until a consensus was reached. 
Accuracy was then computed for (i) LLM baseline solutions, (ii) Human-AI system solutions, and (iii) the top-voted yet unaccepted forum answers. 
The first two metrics reflect experimental performance, while the third serves as a challenging reference baseline. 

\begin{figure*}[t]
	\centering
\includegraphics[width=0.85\linewidth]{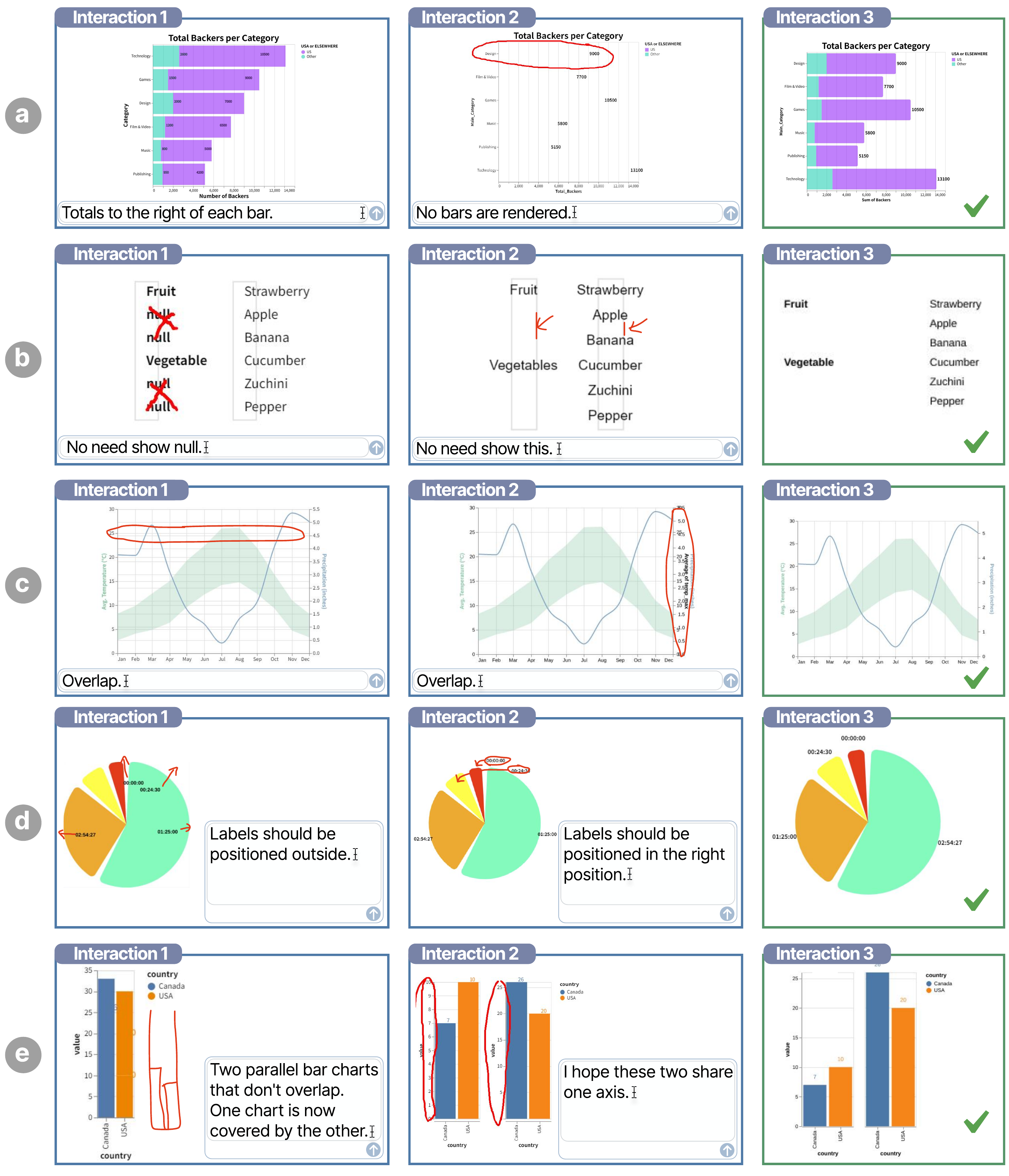}
  \vspace{-1em}
  % \vspace{-2em}
    \caption{Visualization debugging examples during the user study. Participants annotated flawed results and described their intent in natural language. Within one or two rounds of refinement, the system generated corrected visualizations that met their intents.}
\label{fig:System_Example}
    \vspace{-1.em}
\end{figure*}

\stitle{Measures.} 
We define one \textit{round} as a complete cycle of a participant's input followed by a system output of a Vega-Lite specification. 
In the baseline condition, each round corresponded to one interaction with GPT-4o. 
In the system condition, each round was counted whenever the system produced a code preview, including the initial output and subsequent previews after hybrid feedback. Building on this definition, we employed both quantitative and qualitative measures to evaluate accuracy, efficiency, and user experience.

Quantitative measures included: 
(i) \textit{First-turn Accuracy}, defined as whether the initial Vega-Lite specification was correct. In the baseline condition, this was measured after the first round. 
In the system condition, it was measured after the first code preview.  
(ii) \textit{Final Accuracy}, defined as whether the final Vega-Lite specification was correct. In both conditions, this was measured after the last round within the time limit for each case.
(iii) \textit{Subjective ratings}, defined as participants’ self-reported scores on a 5-point Likert scale covering effectiveness, usability, learnability, intention, and satisfaction. These ratings were collected after completing all tasks.

Qualitative measures consisted of semi-structured interviews (15 minutes), which provided insights into user experiences and suggestions to complement the quantitative findings.

\subsection{Examples} 
Figure~\ref{fig:System_Example}(a-e) illustrates representative cases of participants interacting with our system. Users often began by marking flawed results directly on the visualization (\eg highlighting overlaps, missing elements, or misplaced labels) while providing natural language input to describe their intent. Through one or two rounds of iterative interaction, the system was able to refine the specification and generate corrected visualizations that aligned with the users' requirements. This illustrates the advantage of integrating annotation with textual feedback for difficult-to-express debugging issues.

\begin{figure*}[t]
	\centering
\includegraphics[width=\linewidth]{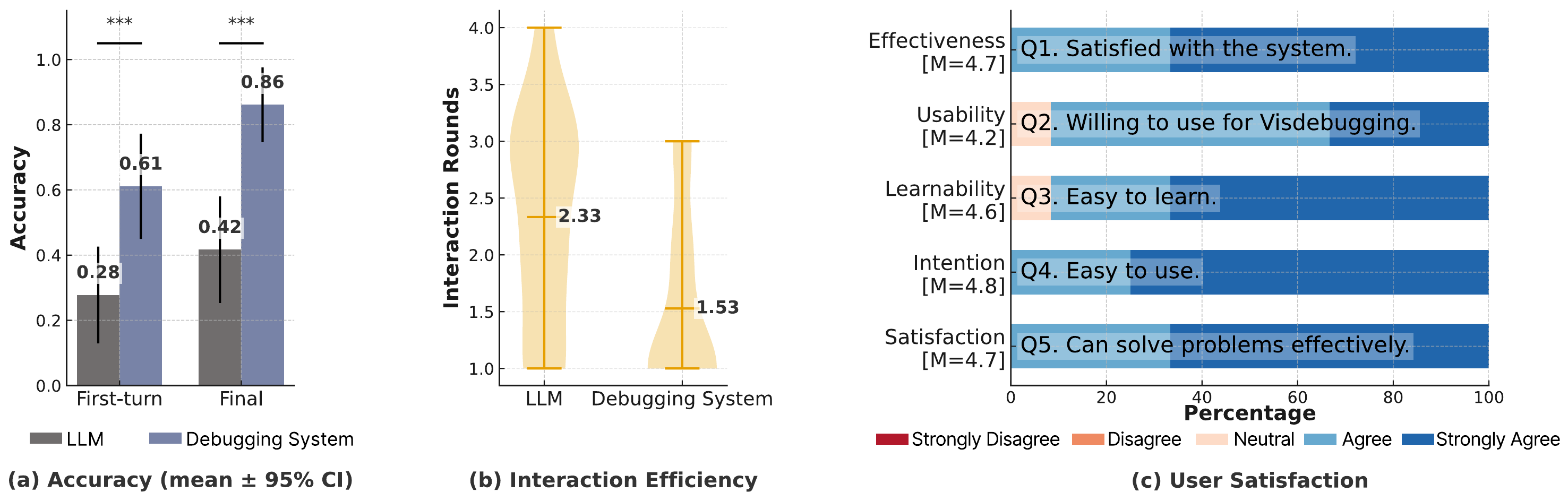}
  % \vspace{-2em}
   \caption{Quantitative and qualitative evaluation results comparing the human-AI collaborative debugging system with the LLM chat-based interface. (a) Accuracy of first-turn and final solutions (mean $\pm$ 95\% CI). (b) Interaction efficiency is measured by debugging rounds to correct solutions. (c) User satisfaction across five dimensions on a 5-point Likert scale.}
\label{fig:accuracy}
    % \vspace{-1em}
\end{figure*}
\subsection{Quantitative Evaluation}
\stitle{System Effectiveness.}
As shown in Figure~\ref{fig:accuracy}(a), first-turn accuracy with our system was much higher than with the LLM baseline (61\% vs.\ 28\%, $p<.001$). This improvement suggests that the combination of multimodal clarification, retrieval augmentation, and code preview provides users with more accurate initial solutions. After iterative rounds of interaction, accuracy further improved to 86\%, surpassing the LLM-chat baseline of 42\%($p<.001$). This highlights the benefit of incorporating hybrid feedback in the loop, which enabled participants to iteratively refine and validate solutions. For reference, forum answers to the same 36 unresolved questions reached 42\% accuracy, underscoring the difficulty of these cases.

\stitle{Interaction Efficiency.}
As shown in Figure~\ref{fig:accuracy}(b), Participants also required significantly fewer rounds to reach a correct solution with our system (M=1.53) than with the baseline (M=2.33, $p<.001$). 
This reduction in interaction rounds indicates that users were able to achieve correct visualizations more efficiently.

\stitle{User Questionnaire.}
As shown in Figure~\ref{fig:accuracy}(c), participants reported high satisfaction with the system (M=4.7) and a strong intention to use it in the future (M=4.8). 
They also rated the system as effective in solving debugging problems (M=4.7). 
Scores for learnability (M=4.6) and usability (M=4.2) were slightly lower relative to the other dimensions. To better contextualize these scores, we next present qualitative insights from semi-structured interviews.

\subsection{Qualitative Evaluation}
The interviews provided deeper context for the quantitative findings, revealing 5 themes about user experiences and challenges.

\uctitle{T1. Clarifying problem intent is helpful for ambiguous posts but redundant for clear cases.}
Participants found the system useful for disambiguating vague or underspecified forum posts (P1, P2, P6).  
Several noted that the tool helped them better understand technical terms (\eg \textit{minEvent}) or tasks where the provided code and problem description diverged (P3, P4, P11).  
However, when the post was already clear, some felt that reviewing auto-generated clarifications introduced unnecessary overhead (P1, P6).

\uctitle{T2. Retrieval augmentation and code preview enable efficient explo\-ration of candidate solutions.}  
Many participants highlighted the value of exploring multiple candidate solutions in parallel, which allowed them to quickly identify promising directions (P3, P4, P10, P12).  
Linked documents were seen as especially useful for deeper exploration (P5). 
As P3 remarked, \textit{ ``The system directly shows me different options, and I can immediately see which ones are feasible.''} At the same time, some users (P11, P12) preferred fewer, higher-confidence results instead of multiple candidates, suggesting that future systems should support adaptive filtering or ranking to balance breadth with precision.

\uctitle{T3. Hybrid Feedback supports intent expression, though it feels unfamiliar to some users.}
Sketching and annotation were considered engaging and helpful for articulating needs that were difficult to describe in natural language (P3, P6, P7, P9, P10, P11).  
Yet, a few participants (P1, P4, P5) felt uncertain about when and how to use multimodal inputs, and preferred direct code modification for simple issues. P4 explained, \textit{``I can highlight key positions through sketches, but for parts that are hard to draw I still prefer to describe them in natural language.''}
Similarly, P5 noted that supporting multiple images for annotation would make the multimodal interaction more practical and flexible. This suggests that to fully realize the potential of hybrid feedback, systems should provide more intuitive affordances to help users decide when and how to apply it.

\uctitle{T4. Large and complex datasets remain a recurring challenge for debugging.} 
Participants consistently reported difficulties when debugging posts that involved very large datasets.  
Strategies included manually simplifying data before inputting it into the system (P1, P6) or wishing for a built-in data simulator to simplify raw data (P2).  
Others encountered expired URLs or inaccessible datasets (P7), which limited their ability to test solutions. These challenges highlight the need for built-in data abstraction or simulation.

\uctitle{T5. Limitations due to problem descriptions and model capability.}  
Participants acknowledged that some failures stemmed not from the system itself but from limitations in the underlying models and problem framing.
Issues such as unclear problem descriptions (P1, P6, P7), hallucinations in generated solutions (P5, P10), and weak model capabilities in spatial reasoning (P11) were cited as barriers.
As P3 noted, \textit{``Some problems are simply beyond what the current models or data allow. The provided code and the actual requirements differ substantially. Achieving the goal requires step-by-step, detailed specification.''}
This suggests that effective debugging requires step-by-step decomposition of complex issues, with humans playing a crucial role in clarifying intent and validating AI outputs.

\stitle{Comparison with General-Purpose AI Assistants.} 
Post-study interviews revealed a preference for our system over general-purpose assistants (\eg Cursor~\cite{Cursor}, Copilot~\cite{Copilot}) due to four key advantages that map to our design modules.
First, regarding intent verification, participants noted that commercial tools function as ``black boxes'' that generate code immediately without verifying user intent (P2, P5, P11). In contrast, our \textit{Multimodal Clarification} module improves transparency by explicitly highlighting its interpretation of image-text inputs. This allows users to correct misunderstandings before generation, reducing the reliance on perfect prompt engineering~\cite{suh2025human}.
Second, regarding knowledge support, users found standard API documentation provided by commercial tools to be often generic or overwhelming (P5, P7). They appreciated that our \textit{Retrieval Augmentation} retrieves targeted, runnable forum examples (P2, P6). This offers contextualized references to justify each fix, rather than leaving users to navigate raw documentation.
Third, regarding solution exploration, commercial tools typically provide single linear fixes, trapping users in trial-and-error loops (P1, P7). Conversely, our \textit{Code Preview} facilitates the parallel presentation of diversified candidate fixes. This enabled participants to compare potential strategies side-by-side without the need for repeated, manual prompting (P8, P10, P12).
Finally, regarding precise refinement, participants highlighted the difficulty of articulating spatial concepts using text alone (P3, P9). Our \textit{Hybrid Feedback} addresses this by enabling visual sketching for spatial constraints. Furthermore, P5 noted that for internal logic errors where sketching is insufficient, the system's ability to integrate error logs and dataflow provides a crucial alternative for precise specification.
\section{Discussion and Future Work}
\label{sec:discussion}
This section summarizes the key insights derived from our study, acknowledges the limitations, and outlines opportunities for integrating the co-debugging system into community platforms.

\subsection{Discussion}
\label{sec:7.1 discussion}
\stitle{Clarifying Problem Intent in Ambiguous Cases.}
Prior work highlights that vague problem descriptions are a major barrier to effective visualization debugging~\cite{Battle2022a,Bostock2011, Luo2025}. Our user study shows that structured clarification helps when posts are ambiguous, but is redundant when problems are already clear. This highlights the importance of adaptive clarification mechanisms that selectively apply follow-up questions only when ambiguity exists (\eg vague terminology, conflicting descriptions).
In addition to textual ambiguity, participants frequently struggled with limited or inaccessible datasets (\eg missing data, oversized data, expired URLs). Such data-related gaps also make problem intent difficult to reproduce and resolve. To address this, debugging systems should support data abstraction, such as privacy-preserving example generators or mock-data simulation~\cite{das2025misvisfix,masson2023chartdetective}, so that representative cases can be shared without exposing sensitive information while ensuring reproducibility of debugging tasks.

\stitle{Adaptive Retrieval Augmentation.} Prior research indicates that over half of LLM-generated programming answers~\cite{kabir2024stack} contain errors. Our experiments show that supplementing models with official documentation, forum posts, and GitHub issues allows retrieval augmentation to provide more reliable solutions. However, not all retrieved solutions remain valid or relevant, since user intent can evolve throughout the debugging process. Incorporating user feedback enables the system to iteratively filter out ineffective solutions and progressively narrow the search space, which may improve both the diversity and the accuracy of the retrieved solutions.

\stitle{Preview for Efficient Comparison.}
Prior work shows that forum-based debugging often involves lengthy trial-and-error exchanges~\cite{oliveira2018exchange,calefato2015mining}.
Our findings suggest that diversified code preview can mitigate this inefficiency by allowing users to compare solutions in parallel. 
However, preferences diverge between breadth and precision. While some users value a wide range of alternatives, others prioritize more targeted results. This suggests that retrieval systems should remain flexible, such as supporting adaptive filtering, consolidating similar code snippets, and lowering the priority of redundant visualizations.

\stitle{Balancing Rich Interaction with Usability.} Multimodal clarification (\eg sketching, annotation, mixed text-visual feedback) can enhance expressiveness~\cite{hong2023conversational,han2023chartllamamultimodalllmchart, shen2024data, WonderFlow}. Our study confirms this potential, but also shows that users were sometimes uncertain when to use these modalities. A practical strategy is to let models propose candidate annotations for user verification~\cite{das2025misvisfix}. For example, a system might infer a refined sketch from a user's prior sketches and then present it for confirmation. To maintain coherent interaction, history memory should be updated when new feedback conflicts with prior intents, ensuring that the evolving context aligns with the user’s current targets.

\stitle{Rethinking Human-AI Role Distribution.}
Prior research indicates that current LLMs face challenges in complex visualization tasks, including hallucinations and reasoning limitations~\cite{kabir2024stack,guo2025deepseek, datadirector}. Our study suggests that while LLMs are effective at proposing useful directions, they often struggle with composition layout and may introduce aesthetic deviation. By contrast, human solutions are more accurate but slower and sometimes suboptimal in complex cases. As models become more proficient at managing complex debugging tasks, human experts may increasingly shift toward supervisory and verification roles rather than directly solving issues~\cite{kabir2024stack}.

\noindent\stitle{Cross-Ecosystem Implications.}
This study focuses on Vega-Lite due to its broad adoption as a high-level declarative grammar~\cite{Satyanarayan2017} and mature ecosystem~\cite{vega_lite_docs}. 
We discuss how these findings apply to broader visualization ecosystems.
For other \textit{declarative grammars} (\eg ggplot2~\cite{Wickham2010}), the shared declarative principle supports transferring our design principles but requires adapting to the decoupling of data transformation and view.
While these grammars prioritize visual encoding, they typically rely on independent upstream data pipelines (\eg pandas~\cite{pandas}) rather than embedded transforms. 
Consequently, effective \textit{Multimodal Clarification} must extend the context beyond the visualization code to capture this external data dependency. 
Additionally, the fragmented structure of visualization specifications makes \textit{Retrieval Augmentation} essential.
Meanwhile, \textit{Hybrid Feedback} conveys aesthetic adjustments that are difficult to capture solely through text.
For \textit{imperative grammars} (\eg D3~\cite{Bostock2011}), the workflow differs by requiring manual management of state transitions (\eg enter-update-exit patterns). 
Since imperative logic often introduces transient errors during interaction or animation (Finding 2.3), \textit{Code Preview} should evolve to dynamic execution monitoring~\cite{hull2023v} by integrating interactive state logging to capture runtime event flows. 
Furthermore, \textit{Hybrid Feedback} should expand beyond static sketches to allow users to verify dynamic behaviors.
Extending beyond programming grammars, our findings also inform the design of \textit{GUI-based visualization tools} (\eg Tableau~\cite{batt2020learning}). 
Current interfaces often constrain user expression by forcing users to manually map complex intents to rigid property panels. 
Our study suggests that \textit{Hybrid Feedback} offers a more direct interaction paradigm where static sketching allows users to visually specify layouts without navigating deep menu trees.

% \add{
% Beyond these ecosystem-specific adaptations, our findings align with research on Code Shaping~\cite{yen2025code}, confirming that \textit{Hybrid Feedback} is a fundamental requirement for effective human-AI collaboration. 
% However, a modality mismatch exists in current approaches as static sketches lack the temporal dimension needed to describe imperative logic, despite their effectiveness in constraining the spatial ambiguity of declarative layouts. 
% Future multimodal systems must therefore evolve beyond static sketching to support dynamic specifications, including state changes or interaction flows, to fully disambiguate user intent across diverse software domains.}
\subsection{Limitations}

Our work also has several limitations:

\stitle{Dataset Scope.} 
To ensure depth in analyzing declarative visualization logic, our study strategically focused on Vega-Lite~\cite{Satyanarayan2017} as a representative high-level declarative grammar. While this focus enables precise error analysis, we acknowledge that specific syntactic error patterns are ecosystem-dependent and differ from imperative libraries like D3~\cite{Bostock2011}. However, as detailed in Section~\ref{sec:7.1 discussion}, our core design framework (\eg Multimodal Clarification, Hybrid Feedback) remains applicable across broader ecosystems. Future work should empirically validate these design principles within diverse programming environments.

\stitle{Benchmark Validity.}
To simulate real-world debugging, our study utilizes web-derived cases from Stack Overflow. We address potential data contamination from Stack Overflow cases through two strategies.
First, the System Evaluation in Section~\ref{sec:Experimental Evaluation} targeted unresolved posts. The absence of accepted solutions prevents ground truth memorization, requiring models to generate novel solutions through reasoning.
Second, the Empirical Study in Section~\ref{sec:human} mitigated memorization through task reformulation. Instead of raw HTML completion, we used structured prompts that explicitly required the ``complete Vega-Lite specification.'' This forces the synthesis of executable code rather than Q\&A.
While these measures minimize reliance on memorization, future benchmarks could further address pre-training risks via data perturbation~\cite{lai2023ds}.

\stitle{Participant Diversity.} Given that the debugging tasks are tightly coupled with code implementation, our recruitment prioritized participants with computer science backgrounds who possess the necessary code comprehension skills. 
However, we acknowledge that this population does not fully represent the broader range of visualization creators, such as data journalists using low-code tools (\eg Tableau~\cite{batt2020learning}), whose debugging strategies may differ. Future work should conduct larger-scale studies with diverse user populations to investigate how mixed-initiative systems integrate into varied real-world workflows.

\subsection{Future Opportunities}
Future opportunities include embedding our co-debugging framework into Q\&A forums, supporting asker-AI task decomposition and human-AI solution selection.

\subsubsection{Asker-AI Agent Collaboration in Task Decomposition.} 

In community forums, askers often provide incomplete or ambiguous problem descriptions. AI agents could first assist with clarification by helping askers refine intent, identify missing details, and ensure that the problem statement is specific enough for effective debugging~\cite{IAI, Instructions}.
Once a clearer problem formulation is established, especially when the gap between a user's initial query and a feasible answer is large, AI agents could propose a structured breakdown of the query into smaller subproblems. These subproblems could be distributed across forum contributors for parallel resolution, leveraging the community's diverse expertise. Finally, the AI agent would synthesize the results into a coherent answer, integrating both human contributions and AI reasoning within the forum workflow.

\subsubsection{Human-AI Agent Collaboration in Solution Selection.}  
Community forums typically rely on practices such as revisions, voting, and reputation~\cite{DBLP:journals/corr/abs-2410-10762,oliveira2018exchange,calefato2015mining} to filter and promote useful answers. Collaborative solution selection could extend the revision process by incorporating AI agents to generate and aggregate multiple candidate solutions, while human agents act as reviewers who provide oversight, feedback, and corrections.  
In addition, voting and reputation systems could be extended to evaluate both human and AI contributions for fairness and reliability. At the user level, reputation scores indicate the credibility of contributions and help prioritize high-quality answers. At the system level, reputation scores can be leveraged to weight candidate solutions from higher-reputation agents more strongly or to use reputation feedback for refining model behavior.  
This integrated process enables iterative refinement loops, where humans and AI jointly converge on high-quality answers, while preserving the transparency and accountability valued in community platforms.
Beyond selecting solutions, forums must also balance meeting immediate user needs with supporting long-term knowledge growth. While AI offers quick solutions, its substantive value resides in producing reusable knowledge structures that align with the long-term goals of Q\&A forums. By supporting adaptable and shareable solutions, AI could contribute both to immediate fixes and to sustainable community knowledge development.

\section{Conclusions}
\label{sec:conclusions}
This paper presented an empirical study of Vega-Lite debugging cases from Stack Overflow, providing a systematic comparison of human and AI assistance in visualization troubleshooting. The analysis revealed a significant gap. Human responses are accurate but slow and inconsistent, whereas AI responses are immediate, yet sometimes misaligned guidance in underspecified or complex cases.  
Building on these insights, we derived design implications for collaborative visualization debugging and developed a mixed-initiative human-AI co-debugging system. The system integrates multimodal clarification, retrieval augmentation, code preview, and hybrid feedback, effectively combining the interpretive strengths of humans with the generative speed of AI.  
A controlled user study validated this approach, demonstrating significant improvements in both accuracy and efficiency compared with forum answers and the LLM baseline. Participants reported that the system was intuitive and particularly effective in handling real-world tasks. 
Our findings highlight the complementary roles of humans and AI in visualization debugging and offer concrete guidance for future collaborative platforms.

\begin{acks}
This paper was supported by the NSF of China (62402409); Youth S\&T Talent Support Programme of Guangdong Provincial Association for Science and Technology (SKXRC2025461); the Young Talent Support Project of Guangzhou Association for Science and Technology (QT-2025-001); Guangdong Basic and Applied Basic Research Foundation (2023A1515110545); Guangzhou Basic and Applied Basic Research Foundation (2025A04J3935); and Guangzhou-HKUST(GZ) Joint Funding Program (2025A03J3714).
\end{acks}

\bibliographystyle{ACM-Reference-Format}
\bibliography{reference}

\clearpage
\appendix
\section*{Appendix}

\renewcommand{\thesubsection}{\Alph{subsection}}

\subsection{Classification and Prompting}
\label{sub:prompting}
In this section, we describe the classifications of questions and operations, as well as the prompt templates used to support our empirical study.

\subsubsection{Question Classification}
\label{subsub:detailedQ1}
Building on prior research in visualization troubleshooting and authoring~\cite{Battle2022a, Bostock2011, rosen2016mobile, Author2023, Wang2024e}, we categorized questions into two main types: troubleshooting issues, which involve identifying and fixing defective visualizations, and non-troubleshooting issues, which encompass authoring issues and system issues~\cite{Battle2022a}.
% Three main question types involve debugging issues, authoring issues, and system issues.

\stitle{Debugging issues} involve flawed visualizations accompanied by embedded code snippets. The primary focus lies in enhancing or rectifying existing visualizations rather than creating new ones from scratch.
\begin{itemize}
    \item \textbf{View.} Focus on improving the aesthetics and design of visualizations. Example: How to get a dashed line in the legend?
    \item \textbf{Data Transformation.} Challenges in processing and restructuring data for visualization. Example: How to encode table-based data?
    \item \textbf{Interaction.} Adding or refining interactive features in visualizations. Example: Is there a way to have a dynamic tooltip in Deneb?
    \item \textbf{Composition and Layout.} Organizing multiple views and combining visualizations into complex layouts. Example: How to create a dashboard with multiple charts?
\end{itemize}

\stitle{Authoring issues} often provide data and seek help in creating visualizations. Example: How do I create a progress bar in \vl?

\stitle{System issues} often post questions tagged with multiple environments or language-related labels pertaining to underlying compatibility limitations. Example: \vl API misbehaving.

The classification was initially generated using GPT-4o and subsequently verified and corrected through manual review to ensure accuracy. 
The prompt template used for implementing this classification is provided in Appendix~\ref{subsub:Details in Prompting}.

\subsubsection{Details in Constructing Code Classification}
\label{subsub:Operation Classification}
\begin{table}[b!]
\begin{center}
\caption{Operation Classification from Vega-Lite with categories and operations.}
\label{tab:operation_classfication}
\vspace{-1em}\small
\begin{tabular}{c|c}
\hline
\textbf{Categories} & \textbf{Operations} \\ \hline
\multirow{19}{*}{\textbf{Data Transformation}} & Data \\ \cline{2-2}
& Aggregate \\ \cline{2-2}
& Bin \\ \cline{2-2}
& Join Aggregate \\ \cline{2-2}
& Stack \\ \cline{2-2}
& Window \\ \cline{2-2}
& Calculate \\ \cline{2-2}
& Density \\ \cline{2-2}
& Extent \\ \cline{2-2}
& Filter \\ \cline{2-2}
& Flatten \\ \cline{2-2}
& Fold \\ \cline{2-2}
& Impute \\ \cline{2-2}
& Loess \\ \cline{2-2}
& Lookup \\ \cline{2-2}
& Pivot \\ \cline{2-2}
& Quantile \\ \cline{2-2}
& Regression \\ \cline{2-2}
& Sample \\ \hline
\multirow{17}{*}{\textbf{View}} & Mark Type \\ \cline{2-2}
& Position Channel \\ \cline{2-2}
& Polar Position Channel \\ \cline{2-2}
& Geographic Position Channel \\ \cline{2-2}
& Text Channel \\ \cline{2-2}
& Hyperlink Channel \\ \cline{2-2}
& Description Channel \\ \cline{2-2}
& Key Channel \\ \cline{2-2}
& Order Channel \\ \cline{2-2}
& Time Unit \\ \cline{2-2}
& Sort \\ \cline{2-2}
& Scale \\ \cline{2-2}
& Legend \\ \cline{2-2}
& Title \\ \cline{2-2}
& Axis \\ \cline{2-2}
& Width/Height \\ \hline
\multirow{4}{*}{\textbf{Interaction}} & Bind \\ \cline{2-2}
& Selection \\ \cline{2-2}
& Interaction \\ \cline{2-2}
& Tooltip \\ \hline
\multirow{5}{*}{\textbf{Composition and Layout}} & Layer \\ \cline{2-2}
& Facet \\ \cline{2-2}
& Concatenate \\ \cline{2-2}
& Repeat \\ \cline{2-2}
& Resolve \\ \hline
\end{tabular}
\end{center}
\vspace{1em}
\end{table}
This section details the process of constructing operation and visualization classifications used in our analysis.

\textbf{Details in Constructing Operation Classification.}
We use classification to categorize key operational codes. To develop an operational classification for analyzing these codes, we reference the Vega-Lite official website, which has been curated and maintained over several years. We summarized the types of operations included in the solutions by using the \textit{Table of Contents} from the \textit{\vl Overview}.
We collected 13 common main properties from the \textit{\vl Overview}, excluded sections that are relatively independent or have content covered in other sections (\eg \textit{Overview, Config, Invalid Data, Property Types}), and organized the remaining content into four main categories: View (including \textit{View Specification, Mark, Encoding, Projection}), Data Transformation (including \textit{Data/Datasets, Transform}), Composition and Layout (including \textit{View Composition}), and Interaction (including \textit{Parameter, Tooltip}). 
We used the section titles from the \textit{\vl Overview} as operation categories. Table~\ref{tab:operation_classfication} summarizes the key operations.

We then applied GPT-4o to label operation types in each best answer and validated the predictions against the official Vega-Lite documentation. 
If a predicted operation was not documented in Vega-Lite, it was corrected during manual review. 
Only validated operations were retained as final labels.
The prompt template used to implement this classification is provided in Appendix~\ref{subsub:Details in Prompting}.

\textbf{Details in Constructing Visualization Type.}
In the \vl Gallery~\cite{vega_lite_docs}, six visualization types are initially provided (Single-View, Composite Mark, Layered Plots, Multi-View Displays, Maps, and Interactive). However, during the labeling process, we observed overlaps among these categories and made the following adjustments: Since examples under Composite Mark can be effectively handled using Layer, we merged it into Layered Plots. Interactive and Multi-View Displays also had significant overlap, particularly in Interactive Multi-View Displays. To resolve the overlap between Interactive and Multi-View Displays, we reassigned cases according to the problem descriptions in the debugging posts.  Additionally, Maps had only 3 cases, making its distribution smaller than other groups, so we incorporated it into Single-View. As a result, we finalized four visualization labels: Single-View Plots, Composite \& Layered Plots, Interactive Charts, and Multi-View Displays.

\subsubsection{Details in Prompting}
\label{subsub:Details in Prompting}
This section introduces the prompt templates used in our empirical study, structured according to the overall workflow of classification and experiments.
The prompting process consists of four main stages: 
(1) classifying the question type, 
(2) identifying key operations, 
(3) generating an initial debugging solution with supplementary resources, 
and (4) refining the solution with compiler feedback. 
We present the detailed templates for each stage below.

We first used the following template to classify each question into categories and subcategories.
\begin{examplebox}{(1) Question Classification Prompt}

\textbf{Input:}
\begin{itemize}
 \item \textbf{Question $Q$:}  
    \begin{itemize}
        \item Title
        \item Description
    \end{itemize}

\item \textbf{Question Classification  $C_1$} (Appendix~\ref{subsub:detailedQ1})
\end{itemize}

\textbf{Output:}
\begin{itemize}
\item The most related question categories.
\end{itemize}
\textbf{Prompt:}
Analyze the following Vega-Lite visualization question and classify it into exactly ONE category with potentially multiple subcategories that best describe the main issues.

Title: \{title\}
Description: \{description\}
Question Classification: \{Question Classification\}

Your response should follow these steps:

1. Choose ONLY ONE category and potentially subcategories that best represent the main issues.

2. Provide a detailed analysis explaining why these specific classifications were chosen.

Please return the response in the following JSON format:
\begin{verbatim}
{
    "category": "Debugging Issues",
    "subcategories": "Visual Style",
    "analysis": "..."
}

\end{verbatim}
\end{examplebox}

We identified the operations involved in each best answer using the following template.

\begin{examplebox}{(2) Operation Classification Prompt}

\textbf{Input:}
\begin{itemize}
    \item \textbf{Question $Q$:}
    \begin{itemize}
        \item Title
        \item Description
    \end{itemize}
    \item \textbf{Best Answer $A$}
    \item \textbf{Operation Classification $C_2$} (Table~\ref{tab:operation_classfication})
\end{itemize}

\textbf{Output:}
\begin{itemize}
\item Up to 3 most significant operation categories.
\end{itemize}

\textbf{Prompt:}
Identify the specific categories, subcategories, and operations involved based on the given classification criteria.

Title: \{Title\}
Description: \{Description\}
Completed Code:
\{Best Answer\}
Operation Classification: \{Operation Classification\}

Please return the response in the following JSON format:
\begin{verbatim}
{
    "category": "category name",
    "subcategory": "subcategory name",
    "operation": "specific operation"
}
\end{verbatim}
\end{examplebox}

We prompted LLMs to generate an initial debugging solution with the template below.

\begin{examplebox}{(3) Visualization Debugging Prompt}

\textbf{Input:}
\begin{itemize}
    \item \textbf{Question $Q$:}
    \begin{itemize}
        \item Title
        \item Description
        \item Images (optional)
    \end{itemize}
    \item \textbf{Supplementary Resources $S$:}
    \begin{itemize}
        \item Documentation (optional)
        \item Examples (optional)
    \end{itemize}
\end{itemize}

\textit{Note: Based on prompt\_type, optional information $S$ includes:}
\begin{itemize}
    \item \textbf{Doc\_only}: Documentation only.
    \item \textbf{Ex\_only}: Example only.
    \item \textbf{Doc\_Ex}: Both documentation and example.
\end{itemize}

\textbf{Output:}
\begin{itemize}
    \item Reasoning Analysis {$A_1$}
    \item Vega-Lite specification {$V_1$}
\end{itemize}

\textbf{Prompt:}
As a visualization expert, please help solve the visualization problem based on the following info:

Title: \{Title\}
Description:  \{Description\}
Supplementary Resources: \{Supplementary Resources\}

Your response should follow these steps:

1. Analyze the user's needs: Understand the core problem and the desired outcome.

2. Provide complete code: Supply the final, complete Vega-Lite specification.

Please return the response in the following JSON format:
\begin{verbatim}
{
    "analysis": "...",
    "vega_lite_spec": {}
}
\end{verbatim}
\end{examplebox}
We optimized the solutions by incorporating compiler feedback, using the following prompt.
\begin{examplebox}{(4) Visualization Debugging Optimization Prompt}

\textbf{Input:}
\begin{itemize}
    \item \textbf{Question $Q$:}
    \begin{itemize}
        \item Title
        \item Description
        \item Images (optional)
    \end{itemize}
    \item First Round Vega-Lite Specification $V_1$
    \item Compiler Feedback $C$:
    \begin{itemize}
        \item Render Status
        \item Logs (optional)
        \item Dataflow (optional)
    \end{itemize}
\end{itemize}

\textit{Note: Based on prompt\_type, compiler feedback $C$ includes:}
\begin{itemize}
    \item \textbf{logs\_only}: Logs only.
    \item \textbf{dataflow\_only}: Dataflow only.
    \item \textbf{logs\_dataflow}: Both logs and dataflow.
\end{itemize}

\textbf{Output:}
\begin{itemize}
    \item Reasoning Analysis {$A_2$}
    \item Optimized Vega-Lite specification {$V_2$}
\end{itemize}

\textbf{Prompt:}
As a visualization expert, please help improve the visualization solution based on the following info:

Title: \{Title\}
Description: \{Description\}
First Round Vega-Lite Specification: \{First Round Vega-Lite Specification\}
Render Status: \{Render Status\}
Logs: \{Logs\}
Data Flow Information:
\{Dataflow\}

Your response should follow these steps:

1. Analyze the user's needs: Understand the core problem and the desired outcome.

2. Provide complete code: Supply the final, complete Vega-Lite specification.

Please return the response in the following JSON format:
\begin{verbatim}
{
    "analysis": "...",
    "vega_lite_spec": {}
}
\end{verbatim}
\end{examplebox}

\subsection{Details of 297 cases}
\label{sub:297 cases}
This section examines the 297 collected cases in detail, focusing on both code complexity and the characteristics of operations.
\subsubsection{Code Complexity}
\label{subsub:4.2.2}
\begin{table}[t]
\centering
\caption{Code Complexity Classification (Q2). Code complexity is classified into four levels: Simple, Medium, Complex, and Extra Complex.}
\label{tab:Code Complexity}
\resizebox{\linewidth}{!}{%
\begin{tabular}{clll}
\hline
\multirow{2}{*}{Type} & \multicolumn{1}{c}{\multirow{2}{*}{Evaluation metric / criteria}} & \multicolumn{2}{c}{Forum Codes} \\ \cline{3-4} 
                      & \multicolumn{1}{c}{}                                              & Question     & Answer   \\ \hline
\multirow{2}{*}{Quantity}              & \# w specs                                                      & 297           & 278      \\
                                                 & \# w/o specs                                                      & 0           & 19      \\ \hline
\multirow{6}{*}{Complexity} & Total \# of keys across specs                               & 12914         & 13606    \\ 
                            & Average \# of keys in a spec                                & 43.48         & 48.94    \\ 
                            & Simple (\#-keys $\leq$ 16)                                      & 38            & 22       \\ 
                            & Medium (\#-keys $\leq$ 24)                                      & 58            & 41       \\ 
                            & Complex (\#-keys $\leq$ 41)                                     & 85            & 85       \\ 
                            & Extra complex (\#-keys $>$ 41)                                  & 116           & 130      \\ \hline
\multirow{3}{*}{Key Differences} & Key decreased                                           & \multicolumn{2}{l}{57}  \\ 
                                 & Key unchanged                                          & \multicolumn{2}{l}{29}  \\ 
                                 & Key increased                                           & \multicolumn{2}{l}{192} \\ \hline
\end{tabular}%
}
\end{table}

To understand whether the solutions provided increase the complexity of \vl, we analyze the code complexity. Table~\ref{tab:Code Complexity} presents our findings on the complexity of question and answer code.

\stitle{Complexity Comparison.}
85 (28.6\%) of the question codes were classified as complex, and 116 (39.1\%) were extra-complex.
In contrast, 85 (28.6\%) of the answer codes were classified as complex, while 130 (43.8\%) were extra-complex.
On average, the complexity increased from 43.48 in the question code to 48.94 in the answer code. However, not all solutions increase complexity. In 20 cases, responders reduced complexity, often by providing more efficient or clearer code, such as using a fold transform to simplify a multi-layered visualization.

\begin{figure}[t!]
	\centering	\includegraphics[width=1\linewidth]{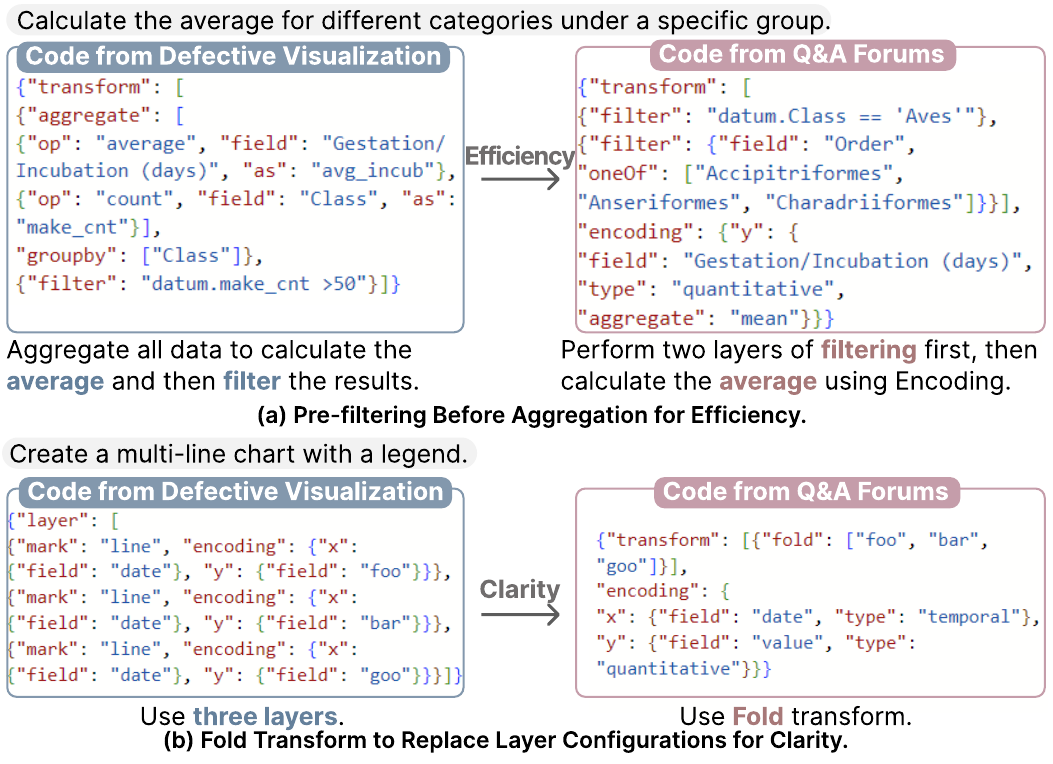}
    % \vspace{-1.5em}
    \caption{Examples of reduced complexity in answer code (Q2). Figure (a) shows a solution that improves efficiency by pre-filtering before aggregation, compared to the user's initial approach of filtering after aggregation. Figure (b) shows a solution that improves clarity by replacing three layers with fold and transform operations.} 
    \label{fig:examples_of_abnormal_code_complexity_patterns}
    % \vspace{-1em}
\end{figure}

\stitle{Cases of Complexity Reduction.}
Not all solutions increase complexity. In 20 cases, responders reduced complexity, often by omitting operations or providing simpler examples. Some cases improved clarity and efficiency. For example, \textit{filtering before encoding averages} reduces redundancy (Figure~\ref{fig:examples_of_abnormal_code_complexity_patterns}(a)), making the code more efficient without changing its functionality. Another example is \textit{using the fold transform} (Figure~\ref{fig:examples_of_abnormal_code_complexity_patterns}(b)), which reduces layers and simplifies the code. These cases demonstrate the responders' ability to provide effective solutions.

\subsubsection{The Characteristics of Operations}
\label{subsub:4.2.3}
% Debugging-related operations refer to the key code modifications made to \vl specifications when solving visualization problems (Figure ~\ref{fig:Q2}). Analyzing these operations helps us understand common methods that forum responses employ to resolve visualization issues.

Analysis of 297 cases reveals key code modifications that forum responses employ to resolve visualization issues.
Operation-level code was categorized into four categories (\textit{Data Transformation, View, Selection \& Interaction, and Composition \& Layout}). Visualization types were classified into four categories (\textit{Single-View Plots, Composite \& Layered Plots, Interactive Charts, and Multi-View Displays}). 

\begin{figure}[t]
	\centering
\includegraphics[width=1\linewidth]{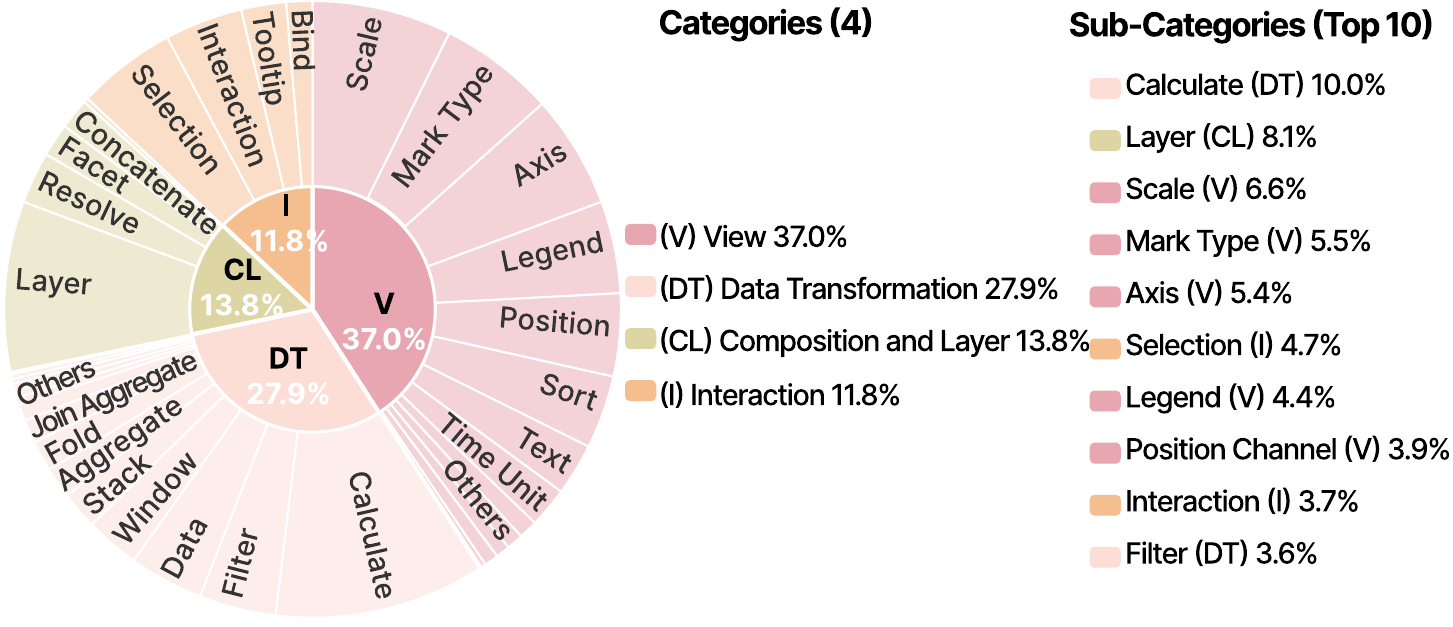}
% \vspace{-1.em}
   \caption{Statistics of the operations (Q2). The inner circle represents four categories, and the outer circle shows individual operations. To ensure clarity, we list the top 10 operation types on the right side. Percentages indicate each operation's occurrence rate among all operations.}
\label{fig:operation_classfication_sunbrust_chart}
    \vspace{-1.em}
\end{figure}

\begin{figure*}[]
	\centering
\includegraphics[width=1\linewidth]{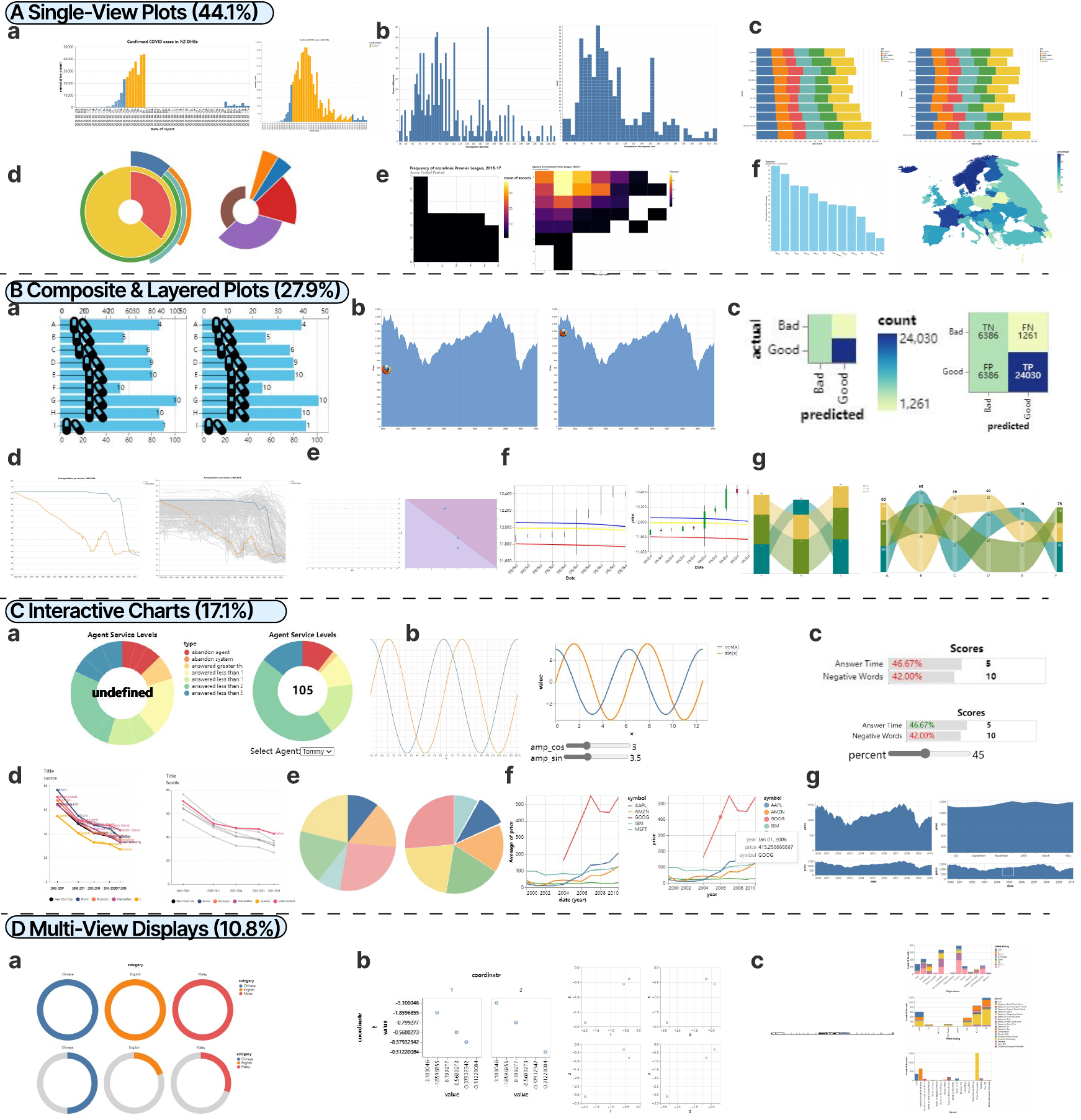}
   \caption{Statistics of the visualization types (Q2). There are four categories, each with examples consisting of two images: one for the problem's visualization and one for the answer's visualization.}
\label{fig:dataset_four_level}
\vspace{-.5em}
\end{figure*}

A comprehensive classification of both visualization types and operation-level code is depicted in Figure~\ref{fig:operation_classfication_sunbrust_chart} and Figure~\ref{fig:dataset_four_level}. We break down the used operation-level code based on their category, accompanied by illustrative examples from visualization types.

\utitle{Operations in View (37.0\%)}: This category is the most commonly used. The most frequent operation is Scale (6.6\%), such as adjusting the color scale in the heatmap shown in Figure~\ref{fig:dataset_four_level} A(e). The second most common operation is \textit{Mark Type} (5.5\%), illustrated by changing the mark from a bar to a map in Figure~\ref{fig:dataset_four_level} A(f). Other complex visualizations include rose charts (Figure~\ref{fig:dataset_four_level} A(d)), box plots (Figure~\ref{fig:dataset_four_level} B(f)), and box-and-whisker plots (Figure~\ref{fig:dataset_four_level} B(g)). \textit{Axis} (5.4\%) are the third most frequent, as shown in Figure~\ref{fig:dataset_four_level} D(b) with dynamic axis modifications. \textit{Position Channel} (3.9\%) is emphasized, as seen in Figure~\ref{fig:dataset_four_level} B(b) and Figure~\ref{fig:dataset_four_level} D(a), where text and icons are aligned in multi-layered visualizations. This indicates that users adjust the positioning of elements within complex visualizations.

\utitle{Operations in Data Transformation (27.9\%)}: This is the second most common type of operation. \textit{Calculate} (10.0\%) is the most frequently used operation. For example, Figure~\ref{fig:dataset_four_level} A(d) illustrates the use of \textit{calculate} to determine the theta of rose charts by computing the difference in cumulative values. \textit{Filter} (3.6\%) is also prevalent, particularly when users need to refine data before visualization. The example in Figure~\ref{fig:dataset_four_level}B(d) involves processing birth rate lines for two countries. 

\utitle{Operations in Composition and Layout (13.8\%)}: This category focuses on the arrangement and alignment of visual elements, particularly in multi-layer visualizations (Layer 8.1\%) (e.g., Figure~\ref{fig:dataset_four_level} B and D). Issues related to composition and layout are often resolved through a combination of operations from the \textit{view} and \textit{data transformation} (e.g., Figure~\ref{fig:dataset_four_level} B(a-g)). Another key operation is \textit{Resolve}, used when users need to share axes or create separate legends between charts, as seen in Figure~\ref{fig:dataset_four_level} D(c).

\utitle{Operations in Interaction (11.8\%)}: Interaction operations set parameters and link them to actions. \textit{Selection} (4.7\%) is common, such as using sliders or dropdown menus to adjust data, axes, or text color (e.g., Figure~\ref{fig:dataset_four_level} C(a-c)). Another aspect of interaction (3.7\%) involves engaging with the visualization. It allows users to highlight or drill down into specific data, both within a single plot (e.g., Figure~\ref{fig:dataset_four_level} C(e)) and across multiple displays (e.g., Figure~\ref{fig:dataset_four_level} C(g)).

% \clearpage
\subsection{Details of 47 cases}
\label{sub: 47 cases}
To evaluate the impact of supplementary resources on LLM performance, we reviewed the forum discussions in 297 cases, identifying 43 cases that included documentation, 2 with examples, and 2 containing both. For cases lacking enough supplementary resources, we manually enriched the dataset by adding relevant guidance from the official \vl documentation or examples from the \vl gallery. When suitable examples were unavailable, we extended our search to GitHub repositories. As a result, a total of 47 cases were supplemented with both documentation and examples. The details of 47 cases are shown in Table~\ref{tab:47_cases_dataset}. 

\begin{table}[t!]
\begin{center}
\caption{47 Selected Cases with Supplementary Resources.} 
\label{tab:47_cases_dataset}
\resizebox{1\columnwidth}{!}{
\begin{tabular}{l|c}
\hline
\multicolumn{1}{c|}{\textbf{Question Types (Operation Details)}} & \textbf{Count} \\ \hline
Color chart with specific colors (Scale, Condition). & 6 \\ 
Add line mark (Rule Properties/Config). & 4 \\ 
Conditional grid format (Axis). & 1 \\ 
Discontinuous issue (Discrete Scales). & 1 \\ 
Hide legend and normalise size (Legends, Circle Properties/Config). & 1 \\ 
Overlay text (Stack, Calculate, Join Aggregate). & 3 \\ 
Text format (Format, Time Unit). & 4 \\ \hline
Data sorting (Window, Calculate, Sort, Order Channel). & 6 \\ 
Data structure (Flatten, Fold, Calculate). & 6 \\ 
Data transformation (Aggregate, Window, Join aggregate). & 3 \\ 
Topn (Window, Filter, Aggregate). & 3 \\ \hline
Create scatter matrix (Pivot, Repeat). & 1 \\ 
Field appears in the wrong chart (Concatenate). & 1 \\ 
Independent color in concatenated plot (Resolve). & 1 \\ 
Share axes in concatenated plot (Resolve). & 2 \\ \hline
Add tooltip (Tooltip Based on Encoding, Tooltip Based on Data Point). & 2 \\ 
Clicking on the main bar chart (Using Parameters Data Extents). & 1 \\ 
Select at a distance (Point Selection Properties). & 1 \\ \hline
\textbf{Overall Question Cases} & \textbf{47} \\ \hline
\end{tabular}

}
\end{center}
\end{table}

% \begin{figure}[]
% 	\centering
% \includegraphics[width=1\linewidth]{figures/deepSeek_case.pdf}
%   % \vspace{-.5em}
%   % \vspace{-2em}
%    \caption{The cases of defective visualization in LLM responses. LLMs failed in zero-shot settings. With additional contextual guidance, some visualizations were corrected, while others remained defective.}
% \label{fig:deepseek_Sample}
%     % \vspace{-1em}
% \end{figure}

% \subsubsection{Characteristics of Fail Cases} 
% \label{subsub: Characteristics of Fail Cases}
% We analyzed uncorrected cases from LLM-generated answers. In Q78603805 (Figure~\ref{fig:deepseek_Sample}), it involved bar highlighting and showed how different models handle the same cases differently. Qwen2.5-72B-Instruct used a simple solution by using red highlighting. In contrast, DeepSeek-R1 took an advanced approach by using an interactive input with color options, but it did not work as expected. While this gave users more flexibility, it also increased the risk of interactive errors, highlighting the trade-off between customization and reliability.
% In Q75013049 (Figure~\ref{fig:deepseek_Sample}), the user wanted to correct a filter function to display only net-zero years. The forum provided examples with 2030 and 2050 as net-zero years, but the model failed to interpret this concept, even though it recognized the filter name as ``Net-zero Target''. This suggests that LLMs still exhibit comprehension biases when handling visualization tasks involving domain-specific knowledge.

\subsection{Index of 36 Debugging Questions without Accepted Answers}
\label{sub:Test Dataset}
This section presents the index of Stack Overflow debugging questions without accepted answers, which served as the test dataset in the user study.

\begin{table}[t]
\centering
\small
\caption{36 Stack Overflow Debugging Question IDs without Accepted Answers.}
\label{tab:question_index}
\vspace{-0.8em}
\begin{tabular}{@{}p{\columnwidth}@{}}
\toprule
\textbf{Question Index}\\
\midrule
76443483, 74860256, 72739971, 79600696, 79714525, 79547333, 79438209, \\
79378627, 79231811, 79201967, 79199739, 77845962, 59685475, 45814950, \\
66624408, 61262606, 72043923, 59134365, 60305836, 57564317, 47140079, \\
74191680, 79386645, 75246451, 79704674, 79485494, 79065168, 79019816, \\
78830816, 78610325, 77860219, 77788976, 78500605, 77992181, 64883154, \\
59729535 \\
\bottomrule
\end{tabular}
\end{table}

% \appendix









\end{document}